\documentclass[12pt,preprint]{aastex}

\shorttitle{Fe-K line probing of AGN with Suzaku}
\shortauthors{Fukazawa et al.}

\begin{document}


\title{Fe-K line probing of material around the AGN central engine with Suzaku}


\author{Yasushi Fukazawa\altaffilmark{1}, Kazuyoshi Hiragi\altaffilmark{1}, Motohiro Mizuno\altaffilmark{1},
Sho Nishino\altaffilmark{1}, Katsuhiro Hayashi\altaffilmark{1}, Tomonori Yamasaki\altaffilmark{1}, 
Hirohisa Shirai\altaffilmark{1}, Hiromitsu Takahashi\altaffilmark{1}}

\affil{Department of Physical Science, Hiroshima University, 1-3-1
Kagamiyama, Higashi-Hiroshima, Hiroshima 739-8526, Japan}
\email{fukazawa@hep01.hepl.hiroshima-u.ac.jp}

\and

\author{Masanori Ohno\altaffilmark{2}}
\affil{Institute of Space and Astronautical Science (ISAS),
Japan Aerospace Exploration Agency (JAXA),
3-1-1 Yoshinodai, Chuo-ku Sagamihara, Kanagawa 252-5120, Japan
}


%\altaffiltext{1}{Visiting Astronomer, Cerro Tololo Inter-American Observatory.
%CTIO is operated by AURA, Inc.\ under contract to the National Science
%\altaffiltext{2}{Society of Fellows, Harvard University.}
%\altaffiltext{3}{present address: Center for Astrophysics,
%    60 Garden Street, Cambridge, MA 02138}
%\altaffiltext{4}{Visiting Programmer, Space Telescope Science Institute}
%\altaffiltext{5}{Patron, Alonso's Bar and Grill}


\begin{abstract}
We systematically analyzed the
high-quality Suzaku data of 88 Seyfert galaxies, about 31\% of which are
Compton-thick AGNs.
We obtained a clear relation between the absorption column density and
the equivalent width (EW) of the 6.4 keV line above 10$^{23}$ cm$^{-2}$,
suggesting a wide-ranging column density of
$10^{23-24.5}$ cm$^{-2}$ with a similar solid and a Fe abundance of
 0.7--1.3 solar for Seyfert 2 galaxies.
The EW of the 6.4 keV line for Seyfert 1 galaxies are typically
40--120 eV, suggesting the existence of Compton-thick matter like the torus
with a column density of $>10^{23}$ cm$^{-2}$ and a solid angle of 
$(0.15-0.4)\times4\pi$, and 
no difference of neutral matter is visible between Seyfert 1 and 2 galaxies.
An absorber with a lower column density of $10^{21-23}$ cm$^{-2}$
 for Compton-thin Seyfert 2 galaxies is suggested to be 
not a torus but an interstellar medium.
These constraints can be understood by the fact that the 6.4 keV 
line intensity ratio against the 10--50 keV flux is almost identical
within a range of 2--3 in many Seyfert galaxies.
Interestingly, objects exist with a low EW, 10--30 eV, of the 6.4
keV line, suggesting that those torus subtends only a small solid
angle of $<0.2\times4\pi$.
Thanks to high quality data with a good signal-to-noise ratio and
the 
accurate continuum determination of Suzaku, ionized Fe-K$\alpha$ emission or
absorption lines are detected from several percents of AGNs. 
Considering the ionization state and equivalent width, emitters and
absorbers of ionized Fe-K lines can be explained by the same origin, and
highly ionized matter is located at the broad line region.
The rapid increase in EW of the ionized Fe-K emission lines at
$N_{\rm H}>10^{23}$ cm$^{-2}$ indicates that the column density of the
ionized material also increases together with that of the cold material.
It is found that these features seem to change for brighter objects with
more than several $10^{44}$ erg s$^{-1}$ 
such that the Fe-K line features become weak. This extends the previously
known X-ray Baldwin effect on the neutral Fe-K$\alpha$ line to ionized
emission or absorption lines. 
The luminosity-dependence of these
properties, regardless of the scatter of black hole mass by two orders of
magnitudes, indicates that the ionized material is associated with the
structure of the 
parent galaxy rather than the outflow from the nucleus.
\end{abstract}


\keywords{galaxies: active --- galaxies: Seyfert --- X-rays: galaxies}



\section{Introduction}

X-ray spectra from Seyfert galaxies are never represented by a simple
power-law, but a wide range of reprocessed features such
as absorption, emission and absorption lines, reflection, and so on
(Turner and Miller 2009).
Such features are vital to probe the surrounding material
around central supermassive black holes, such as accretion disk or
flow, torus, and clouds.
Since such materials are fuel for massive black holes, 
information on the evolution of the supermassive black holes
can be obtained by studying them.

The most prominent feature is the photoelectric absorption of the
continuum in the soft X-ray band, where both cold and warm absorbers are
found.
The cold matter is thought to be associated with a molecular torus.
For Compton-thin objects with a cold absorption column density of
$N_{\rm H}<10^{23}$  cm$^{-2}$, $N_{\rm H}$ can be 
determined by observations below 10 keV.
Based on the observations with BeppoSAX, Swift/BAT, and INTEGRAL, a
significant fraction of Seyfert galaxies exhibit a Compton-thick cold
absorption (Risaliti et al. 1999; Beckmann et al. 2006; Tueller et
al. 2008), and the direct nuclear X-ray emission of such objects 
cannot be observed below 10 keV.
The location and geometry of the cold absorber is now extensively studied.
Matt (2000) pointed out that the Compton-thin absorber 
differs from the Compton-thick molecular torus and is
associated with the interstellar medium in the parent galaxy.
The rapid time variation of the absorption column density indicates that some
of the cold absorbers exist as a blob-like cloud (e.g. Elvis et al. 2004).
Compton-thick Seyfert 2 galaxies also exhibit a complex absorption
feature, possibly due to partial covering and so on (Comastri et al. 2010).
Some of the Compton-thick objects exhibit no optical activity and weak
scattering X-ray component, indicating that the torus is geometrically
thicker than ever thought (Ueda et al. 2007).
A warm absorber due to ionized clouds has often emerged.
Although initially recognized through an ionized absorption edge in the soft
X-ray band (Halpern 1984), 
a lower-density absorber has also recently emerged
via detection of soft X-ray and UV absorption lines (e.g. Kaspi et al. 2002).
These absorption materials are also observed as reflectors or
scatterers, and 
many emission lines in the optical to X-ray band are detected from
Seyfert galaxies (e.g. Kaspi et al. 2002), as are wide-ranging 
ionization states; for example, the
ionization state of iron ranges from 0 to 26.

These materials are considered to be related to the evolution of the
central black hole and engine.
A recent X-ray survey revealed that heavily absorbed AGNs lack
high luminosity (e.g. Ueda et al. 2003; Winter et al. 2009),
while blue-shifted absorption lines sometimes emerge in high
luminosity AGNs (Pounds et al. 2003; Reeves et al. 2003), 
indicating a relativistic massive outflow, which is
promising feedback for the parent galaxy evolution.
With this in mind, it is important to obtain a general view of materials
around the supermassive black hole to understand the
coevolution of the black hole and galaxy.
The physical properties of environmental materials are subject to the
luminosity of the central engine.
It is also important to understand the complex X-ray spectra of
Seyfert galaxies and derive the intrinsic spectral shape, 
to clarify the physical view of the central engine,
while correct modeling of the X-ray continuum emission is also required
to evaluate the disk line and the reflection component.

However, the absorption column density and intrinsic luminosity of
Compton-thick Seyfert galaxies cannot be measured below 10 keV.
In addition, reflection component constraint requires 
wide X-ray band spectroscopy.
BeppoSAX and/or XMM-Newton has provided opportunities for such studies,
but the energy resolution of BeppoSAX around Fe-K lines is too poor to
resolve the broad/narrow, neutral/ionized, or absorption
lines respectively.
The variability of Seyfert galaxies means broadband X-ray 
spectral shapes cannot be ensured 
when observations below and above 10 keV are not simultaneous.
However, the Suzaku XIS/HXD combination (Mitsuda et al. 2007; Koyama et
al. 2007; Takahashi et al. 2007; Kokubun et al. 2007)
is relatively powerful for such studies,
thanks to its wide X-ray band, good energy
resolution, and well-calibrated response.
Since the typical exposure time of AGNs with Suzaku is longer
than 50 ks, the signal-to-noise ratio of data is very high.
Here, we report systematic studies of the Fe-K line features of
Seyfert galaxies and their dependence on intrinsic luminosity and
the cold absorber.
We evaluated the Fe line features as free from the modeling as possible
to avoid the model dependence of physical quantities, such as 
solid angles and Fe abundance.
The disk line was not treated in this paper, since it depends on 
continuum modeling.

Throughout this paper, we adopt a Hubble constant of $H_0=$70 
km s$^{-1}$ Mpc$^{-1}$.
The solar abundance ratio is referred to as the
solar photospheric values of 
Anders and Grevesse (1989) for photoelectric absorption and reflection, 
and the cross-section for the absorption model is set to that of 
Ba\l uci${\rm \acute{n}}$ska-Church and McCammon (1992).
The errors are typically shown at a 1$\sigma$ confidence level for one
interesting parameter.

\section{Data Sample}

We selected Seyfert galaxies from the archived Suzaku data as of
September 2009.
Initially, we examine XIS and HXD spectra, whereupon 
the objects detected with HXD-PIN were chosen.
The HXD-PIN detection is important to constrain the Compton-thick
absorption column density and measure the intrinsic luminosity free
from absorption.
This condition leads to the selection of objects which are bright enough
to analyze the Fe-K line feature with a good signal-to-noise ratio.
Consequently, we analyzed 88 objects, about 36\% of which were Seyfert 1
galaxies ($N_{\rm H}<10^{22}$ cm$^{-2}$), about 33\% Compton-thin
Seyfert 2 galaxies 
($10^{22}$ cm$^{-2}$ $<N_{\rm H}<5\times10^{23}$ cm$^{-2}$), and about
31\% Compton-thick Seyfert 2 galaxies 
($N_{\rm H}>=5\times10^{23}$ cm$^{-2}$).
All the objects were observed in XIS 5x5 or 3x3 modes and 
normal HXD mode.
We screened the data with Suzaku standard selection criteria and the
XIS photons accumulated within 4 arcmin of the object, with 
the XIS- 0, 2, and 3 data coadded to derive the spectrum.
The XIS rmf and arf files were created with {\tt xisrmfgen} and {\tt
xisarfgen} (Ishisaki et al. 2008), respectively, and
the XIS detector background was estimated with {\tt xisnxbgen} (Tawa et
al. 2008).
For the HXD, the "tuned" PIN and GSO background was used (Fukazawa
et al. 2009) and
the good time interval (GTi) was determined by taking the logical-and
of GTIs among XIS data, HXD data, and HXD background data.
For the XIS and HXD-PIN, 
CXB was added to the background spectrum thus obtained, although
negligible for the HXD-GSO.
The GSO response correction file 
{\it ae\_hxd\_gsoxinom\_20070424.arf} was also utilized, which compensated
for any disparity of 10--20\% between the Crab spectral model 
and the data for the current response matrix (Takahashi et al. 2008).
In the simultaneous spectral fitting of XIS and HXD, 
a constant factor of 1.13 was introduced for both PIN and GSO against the XIS.

\section{Analysis of the Fe-K line features}

In order to study the Fe-K line features, we initially determined the
baseline continuum model.
Since the Suzaku spectra of most Seyfert galaxies cannot be expressed by
a simple
power-law model plus absorption, we added a reflection component and
an Fe-K line and fitted the spectra above the 3 keV (model A).
We applied the {\it pexrav} model (Zdziarski et al. 1995) 
for the reflection component.
We include both the Galactic and object-intrinsic
absorptions.
The former was fixed to the value in Dickey \& Lockman (1990).
For the latter, the column
density and Fe abundance were left free and other elemental abundances
were fixed to 1 solar.
For the reflection component,
Fe abundance and reflection fraction $R$ were left free, 
but the Fe abundance was tied to that of the absorption.
The parameters of the input power-law emission for the pexrav model were
fixed to that of the above
power-law model, and the inclination $\theta$ to
$\cos\theta=0.207$.
$\theta$ was related to the relative intensity of the reflection
component, alongside $R$. 
Since the latter was a free parameter, the fixing of $\theta$
did not significantly affect the results on continuum modeling.
After obtaining these spectral parameters, we restricted the energy band
to the range 5--9 keV, 
and fixed the absorption, reflection fraction, and powerlaw
photon index, where the powerlaw normalization was the single free
parameter for the continuum model.
We included five Gaussians (model B), 
considering lines of 6.4 keV, 6.7 keV (He-like), 7.0
keV (H-like), and 7.09 keV (neutral Fe-K$\beta$). 
Although the line energy and width of 6.4 keV were let free. 
the center energy of other lines was
fixed to the value in the rest frame, and their line width was 0 keV.
while a negative value was allowed for the normalization of ionized Fe-K
lines; 6.7 and 7.0 keV.
The intensity of the Fe-K$\beta$ line was fixed to 0.125 of the
Fe-K$\alpha$ intensity (Palmeri et al .2003),
since it is usually too weak to measure accurately in most cases.
We do not consider a disk-line, since
the disk-line feature could affect the continuum modeling, leading to
the systematic uncertainty of the equivalent width. 
However, since its shape resembles that of the reflection continuum, 
we believe that the affection is reduced by including the {\it pexrav}
model even if the disk-line exists. 

Apart from the line studies, we also obtained the absorption column density
for discussion.
Although the soft X-ray band could not be reproduced well by a simple model,
we fit the spectra with a single power-law model plus Gaussian,
multiplied by the absorption and 
differing from model A in that the metal abundances of all elements
were fixed to 1 solar (model C).
The energy band for fitting was limited to higher energy than for
model A so that complex spectral features in the soft X-ray band could 
be neglected.
Recently, it was revealed that the absorption feature of Seyfert
galaxies is more complex than is represented by the single column
density (e.g. Comastri et al. 2010),
meaning that the column density obtained in this fitting is a typical or
representative value for each object.
Since we excluded the soft X-ray band, 
the complex absorption feature in the soft X-ray band did not affect the
results for the Fe-K line intensities and the flux in 10-50 keV.

Finally, in order to obtain the Fe-K edge properties, we added the edge
model into model B. 
Since the reflection and absorption include a Fe-K edge
feature, the Fe abundance of these components was fixed to 0 solar and
other elemental abundances to 1 solar, respectively.
We derived the edge energy by allowing the edge energy and depth to be free.
To obtain the edge depth $\tau$, the edge energy was fixed to 7.112
keV (X-ray Transition Energy Database {\tt http://www.nist.gov/physlab/data/xraytrans/index.cfm}).

\begin{figure}
%\plottwo{07062612_702117010_fesp.eps}{06072112_701036010_fesp.eps}
%\plottwo{07111721_701035020_fesp.eps}{08111202_703022030_fesp.eps}
%\plottwo{07070305_702005010_fesp.eps}{06080816_701030010_fesp.eps}
%\plottwo{06102205_701029010_fesp.eps}{06120913_701021010_fesp.eps}
\epsscale{.80}
\plottwo{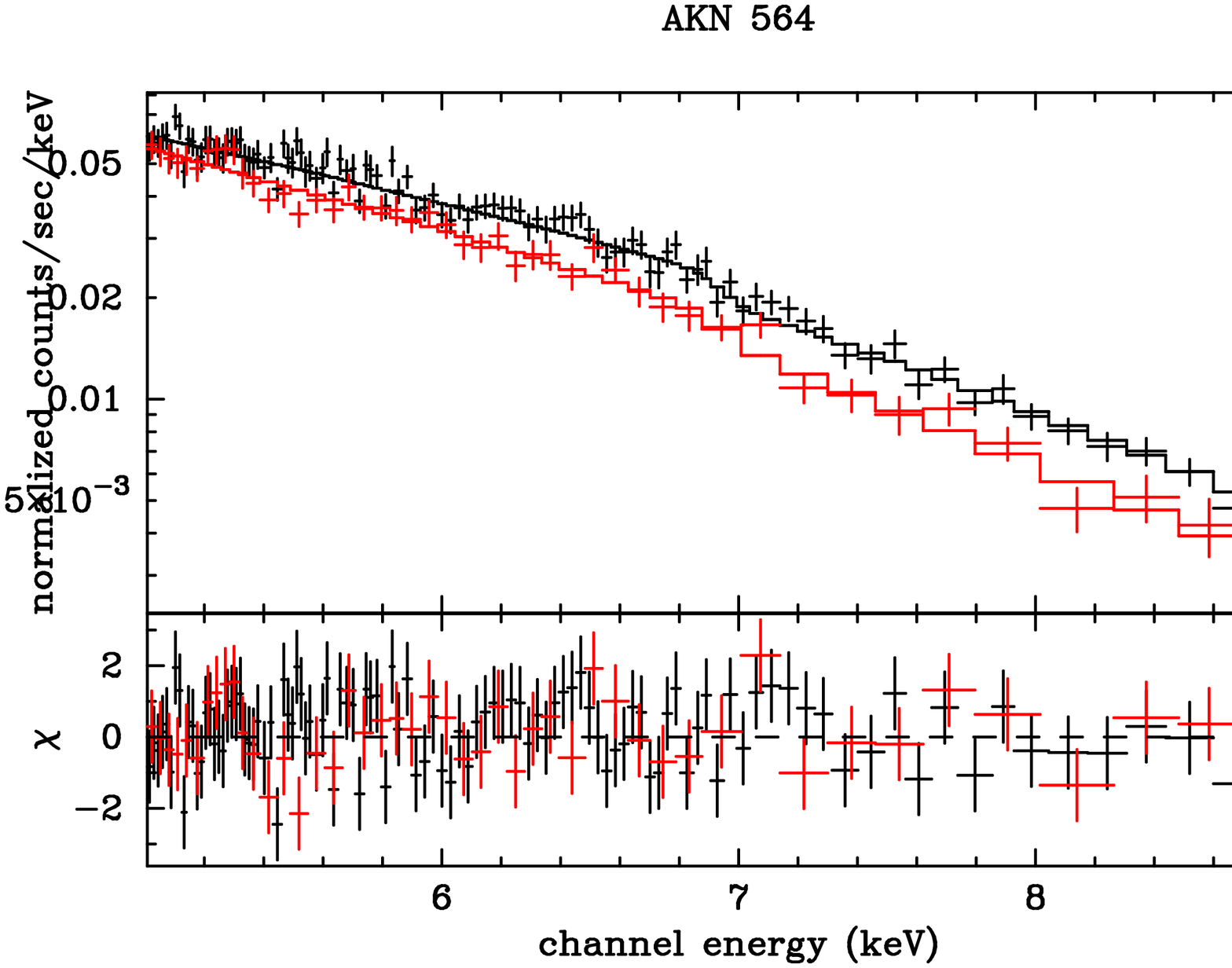}{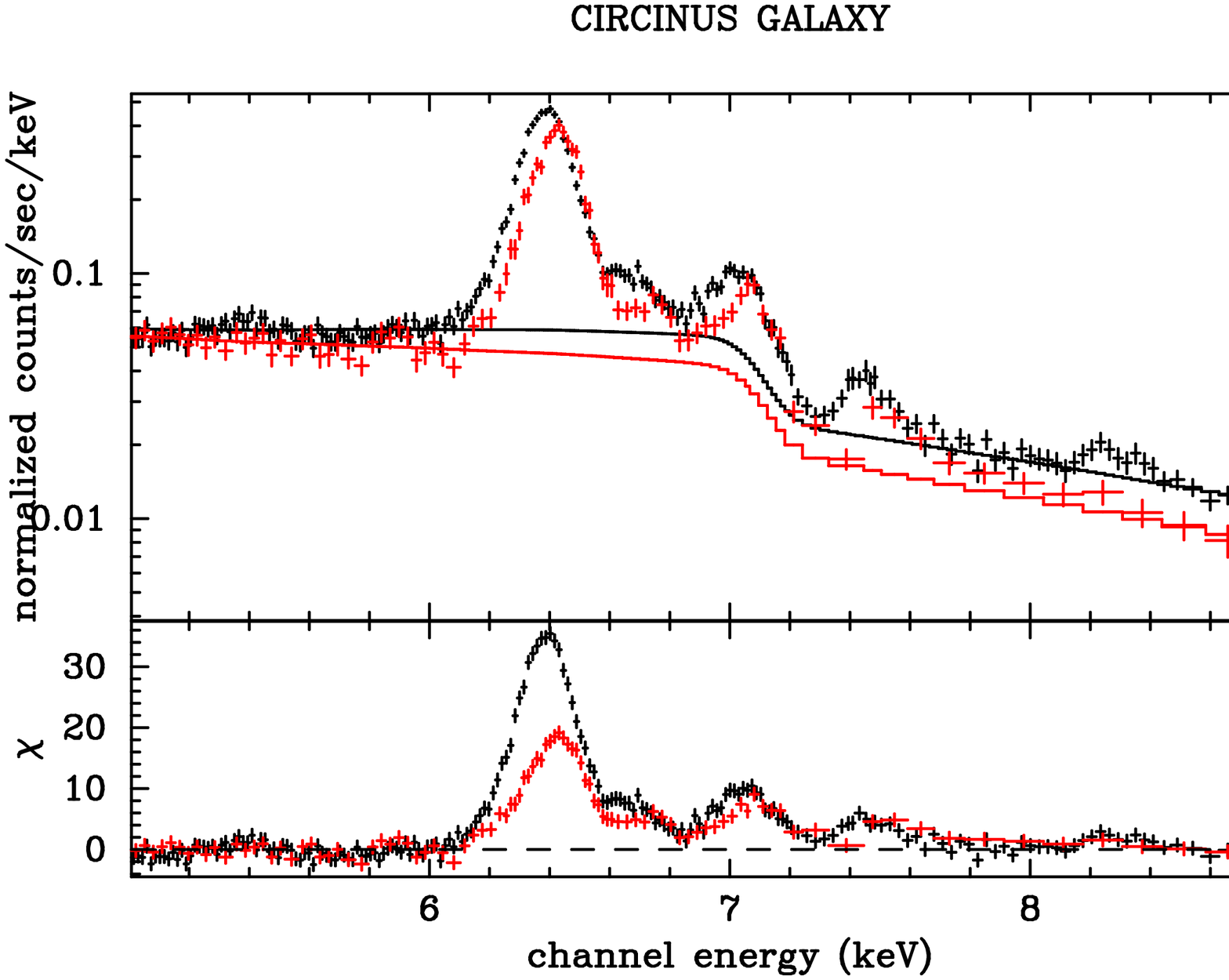}
\plottwo{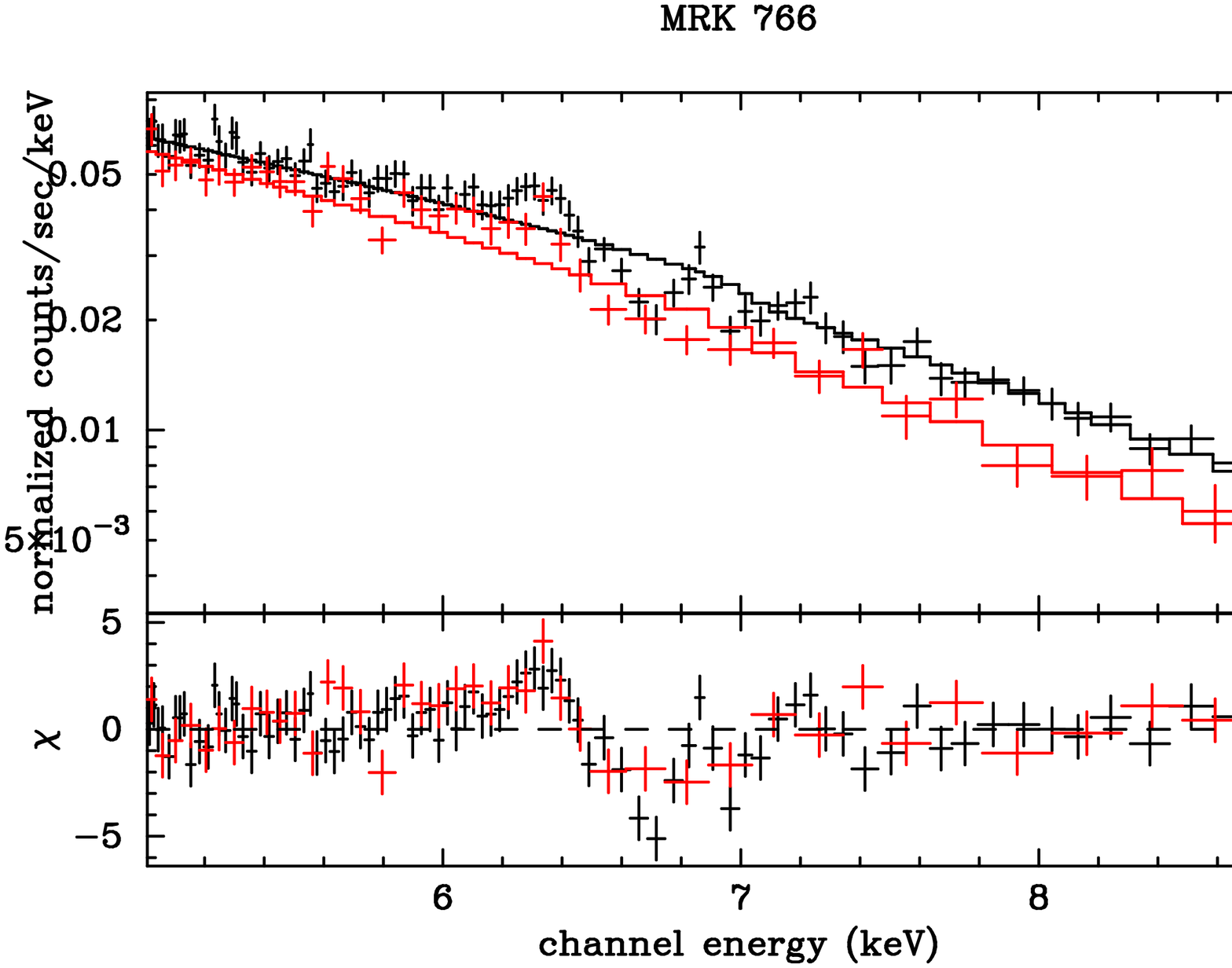}{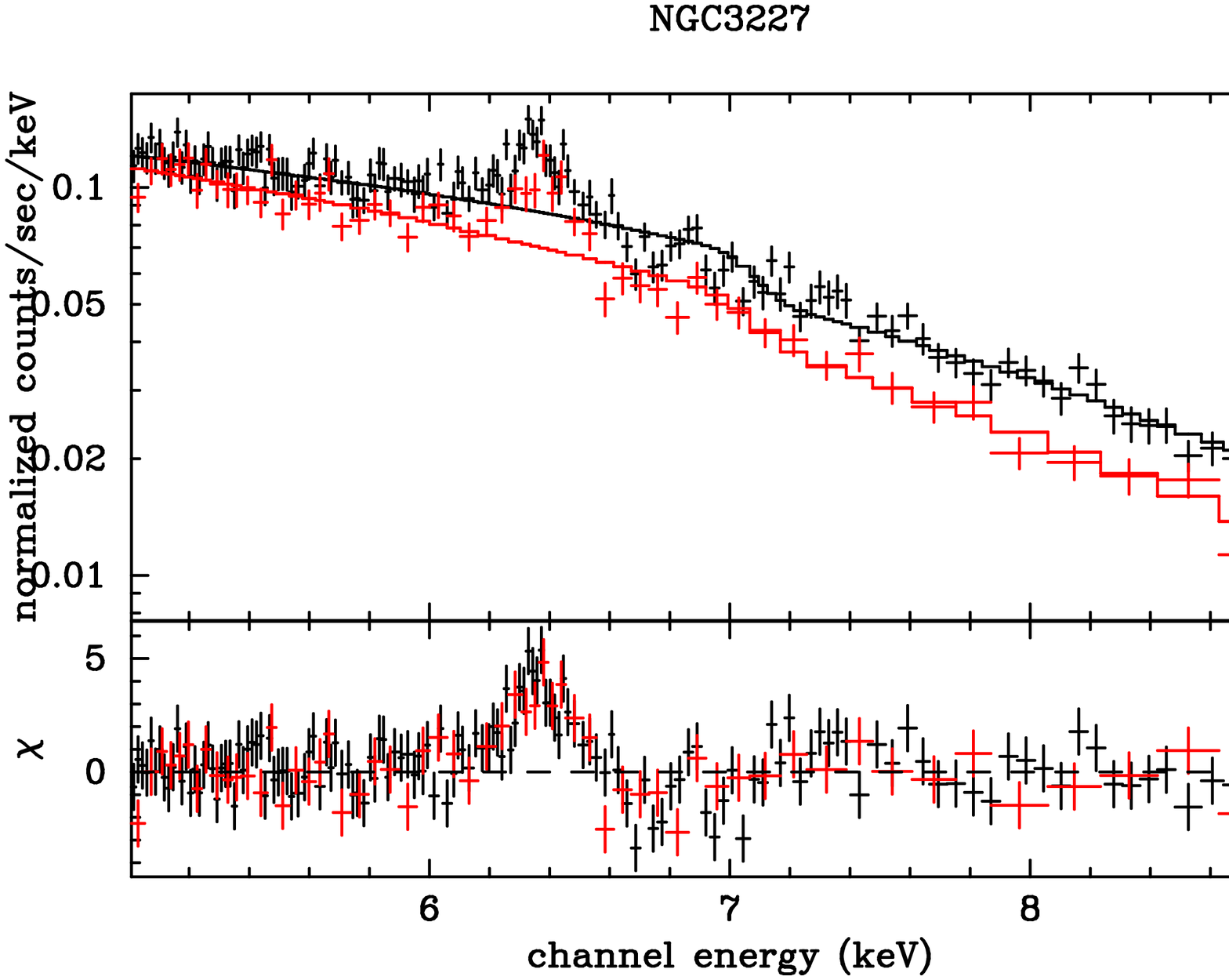}
\plottwo{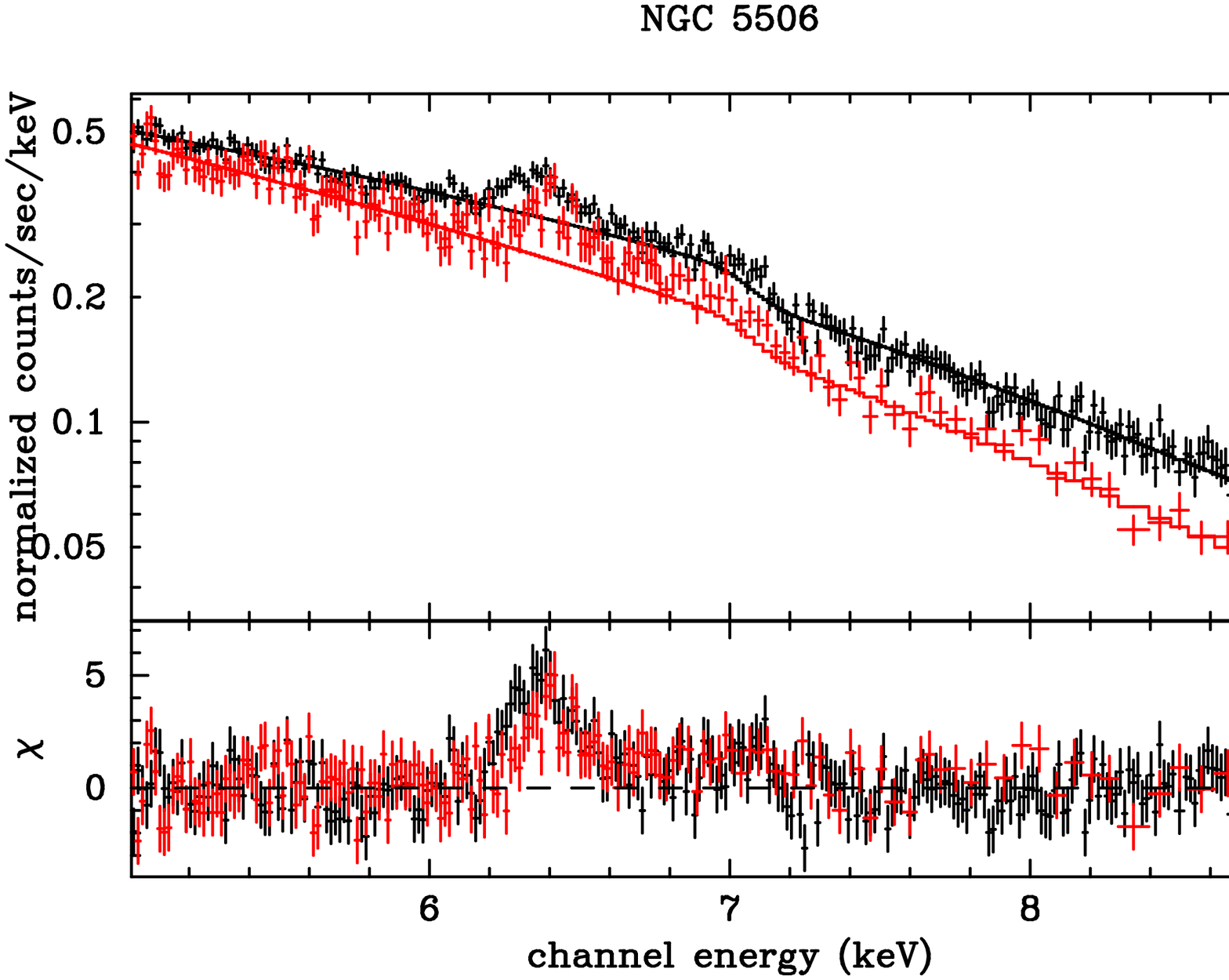}{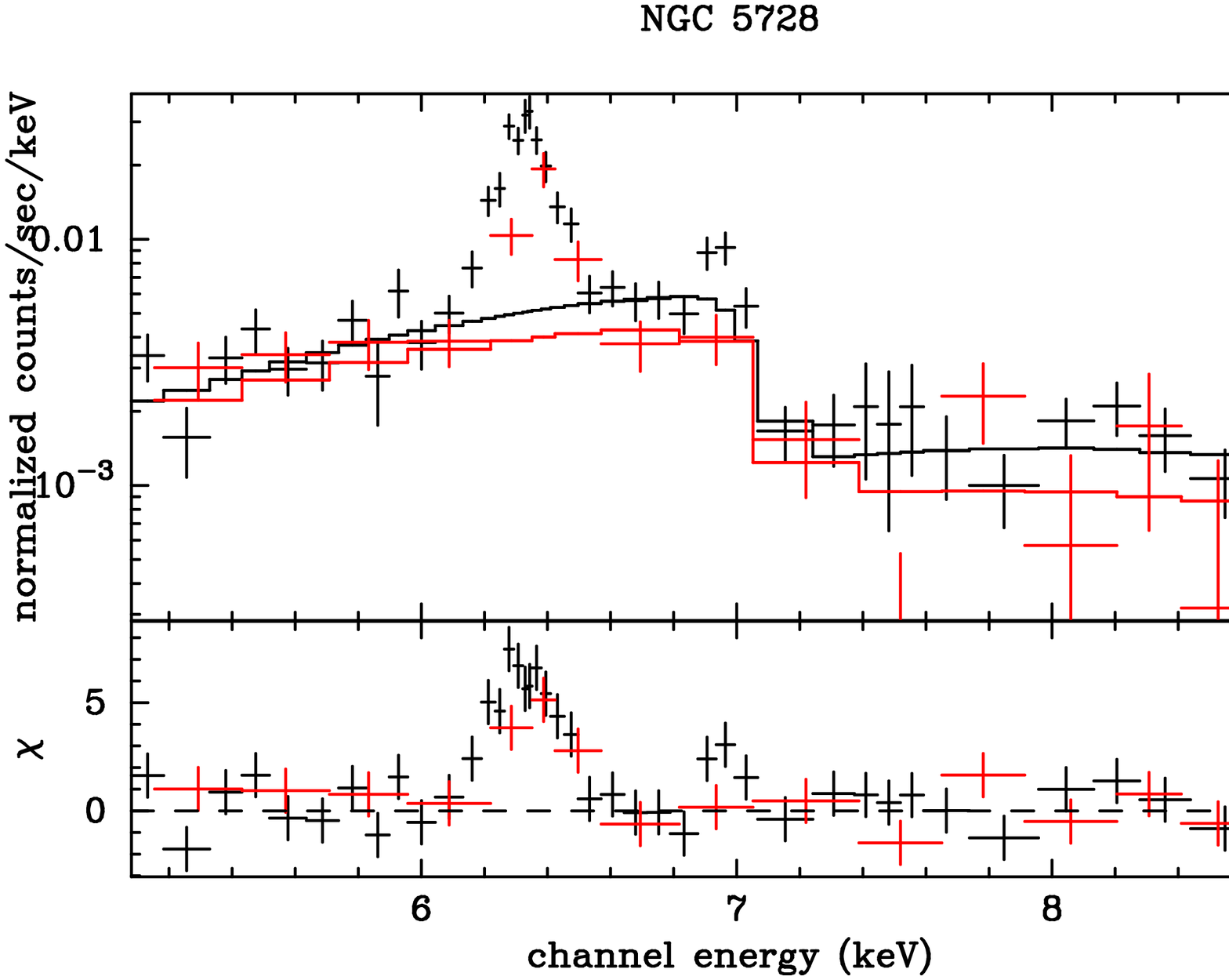}
\plottwo{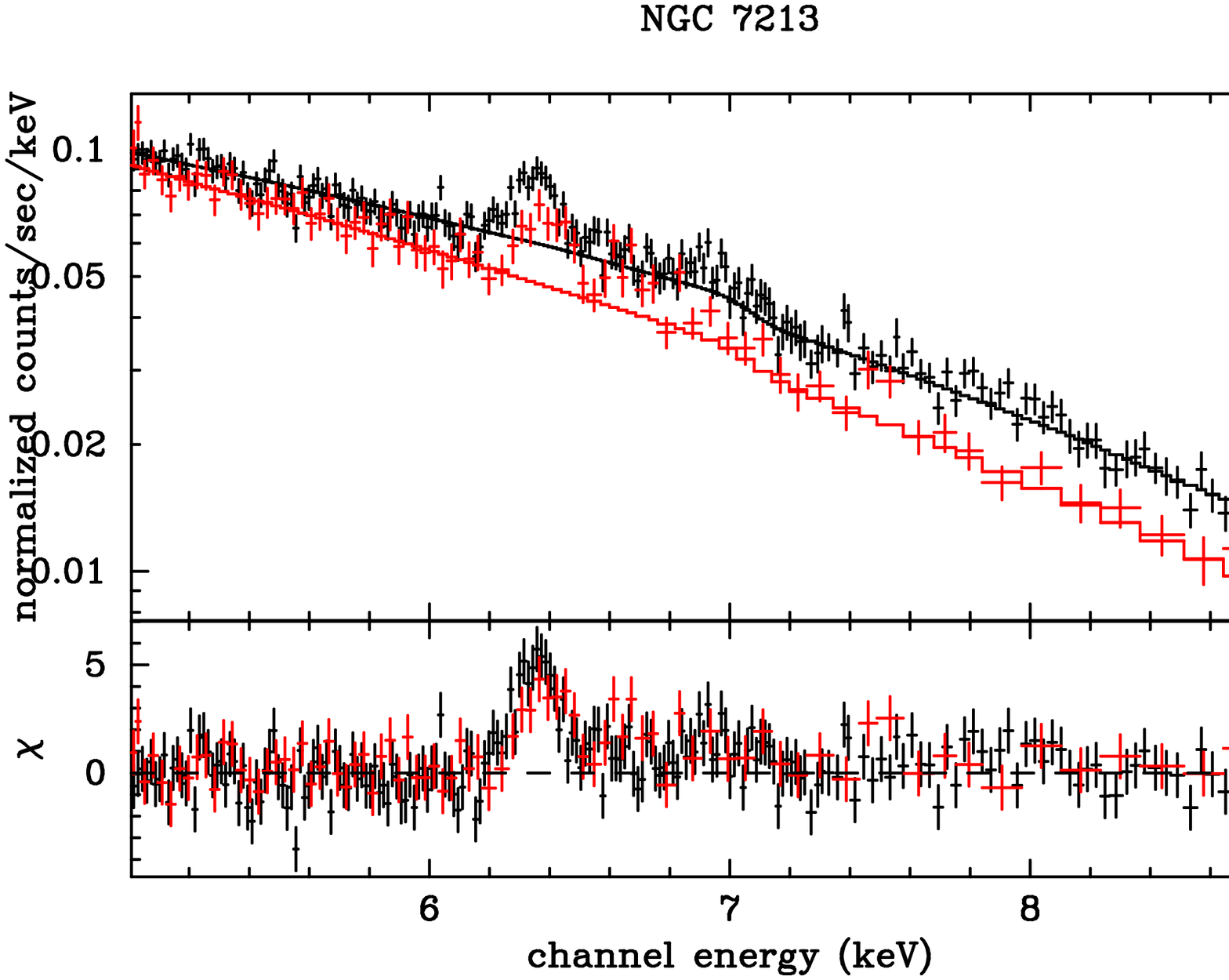}{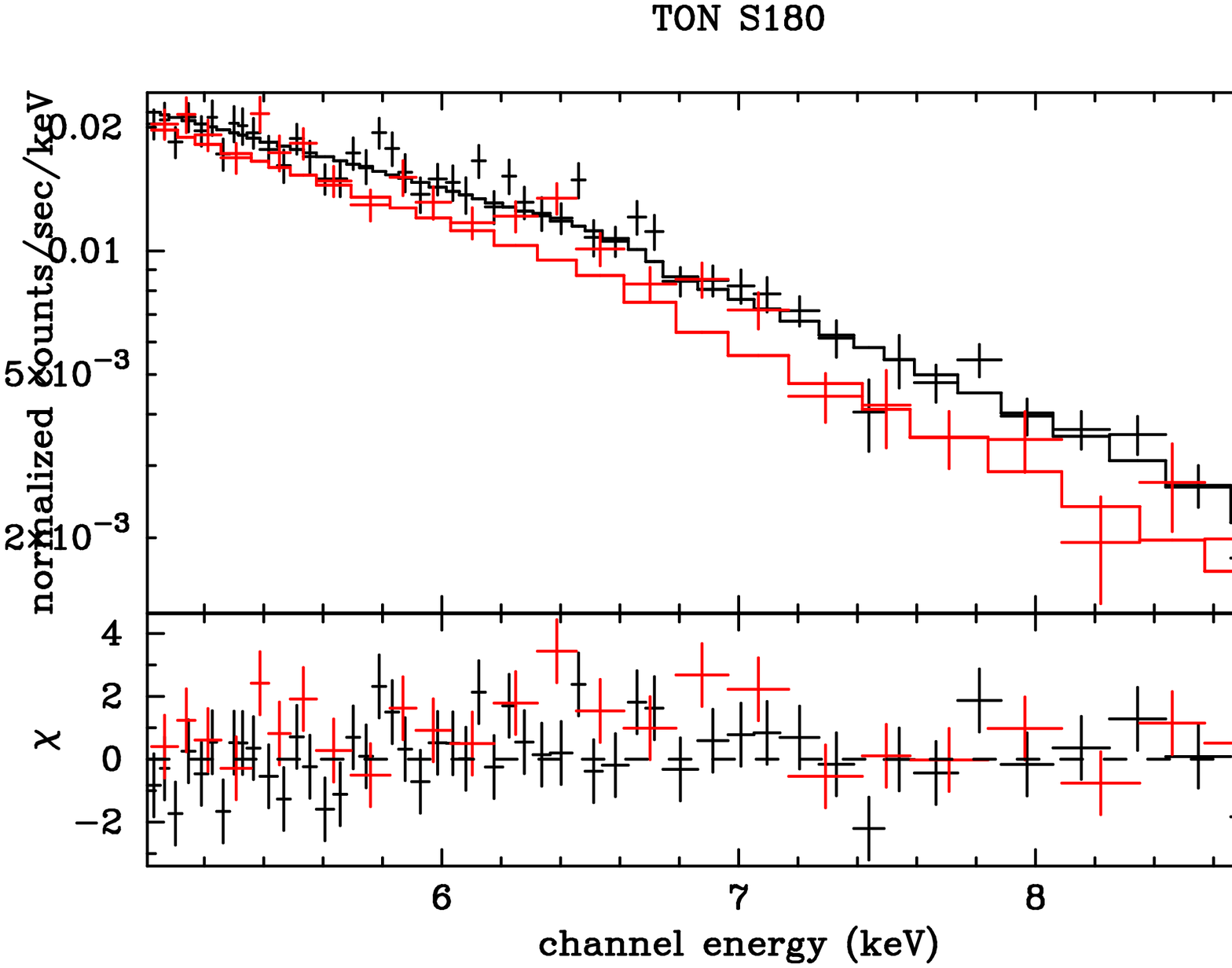}
\caption{ Example of the XIS spectra around the Fe-K features. From top
 to bottom: ARK 564, CIRCINUS GALAXY, MRK 766 (701035020), NGC 3227
 (703022030), NGC 5506 (701030010), NGC 5728, NGC 7213, TON S180. The solid
 line in the top panel is the best-fit continuum model. The bottom panel
 shows the residual. Black and Red correspond to the XIS-F and XIS-B,
 respectively. \label{fespec}}
\end{figure}

In table 1, we summarize the fitting results for Fe-K line features, together
with the absorption column density, X-ray flux, and luminosity in 10--50 keV
where the absorption less affects the observed flux: for example, the
observed flux is a maximum of 1.5 times as small as the absorption-corrected
one in the case of $N_{\rm H}=2.5\times10^{24}$ cm$^{-2}$ with 1 solar
abundance.
Figure \ref{fespec} shows a sample of XIS spectra around the Fe-K line
features, where only the best-fit continuum model is shown for comparison.

\section{Results of the Fe-K Line Features}

\subsection{Neutral Fe-K Lines and Edge}

\begin{figure}
\epsscale{1.0}
\plotone{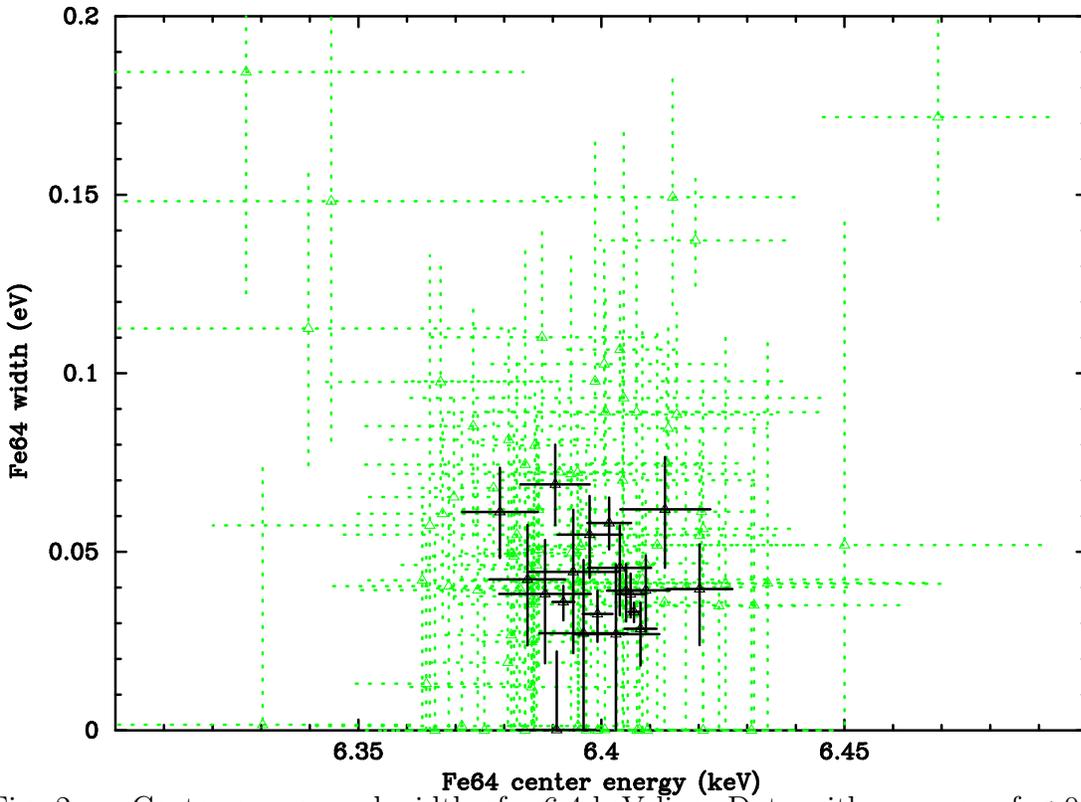}%fe64enesig.eps
\caption{ Center energy and width of a 6.4 keV line. Data with an error of 
 $<0.01$ keV for the line energy are denoted as solid error bars.
\label{fe64enesig}}
\end{figure}

Figure \ref{fe64enesig} shows the plots of center energy and 
width for the 6.4 keV line.
Objects with well-determined values reveal center energy within 
6.37--6.42 keV and 
average and variance of 6.399 and 0.008 keV, respectively.
Although the uncertainty of the XIS energy scale is 0.01--0.03 keV at 6 keV, 
the average center energy of the 6.4 keV line is consistent with a
neutral or low-ionized Fe-K fluorescence line, which
constrains the ionization state to $<12$ (Kallman et al. 2004).
The line energy is not dependent on the absorption column density;
the average and variance are 
6.390 and 0.017 keV for $N_{\rm H}<5\times10^{21}$ cm$^{-2}$, 
6.397 and 0.009 keV for $5\times10^{21}<N_{\rm H}<1\times10^{23}$ cm$^{-2}$, and
6.400 and 0.014 keV for $N_{\rm H}>1\times10^{23}$ cm$^{-2}$, respectively.
The width is constrained to be $<3300$ km s$^{-1}$ 
or $<70$ eV for objects with well-determined values, 
and also independent of the absorption column density;
the average and variance are 
55 and 25 eV for $N_{\rm H}<1\times10^{22}$ cm$^{-2}$, and 
38 and 26 eV for $N_{\rm H}>1\times10^{22}$ cm$^{-2}$.
These constraints are almost similar to those derived for certain individual
objects (e.g. Yaqoob et al. 2007; Shirai et al. 2008; Awaki et
al. 2008), but the results confirm this as a common feature.

\begin{figure} 
\plotone{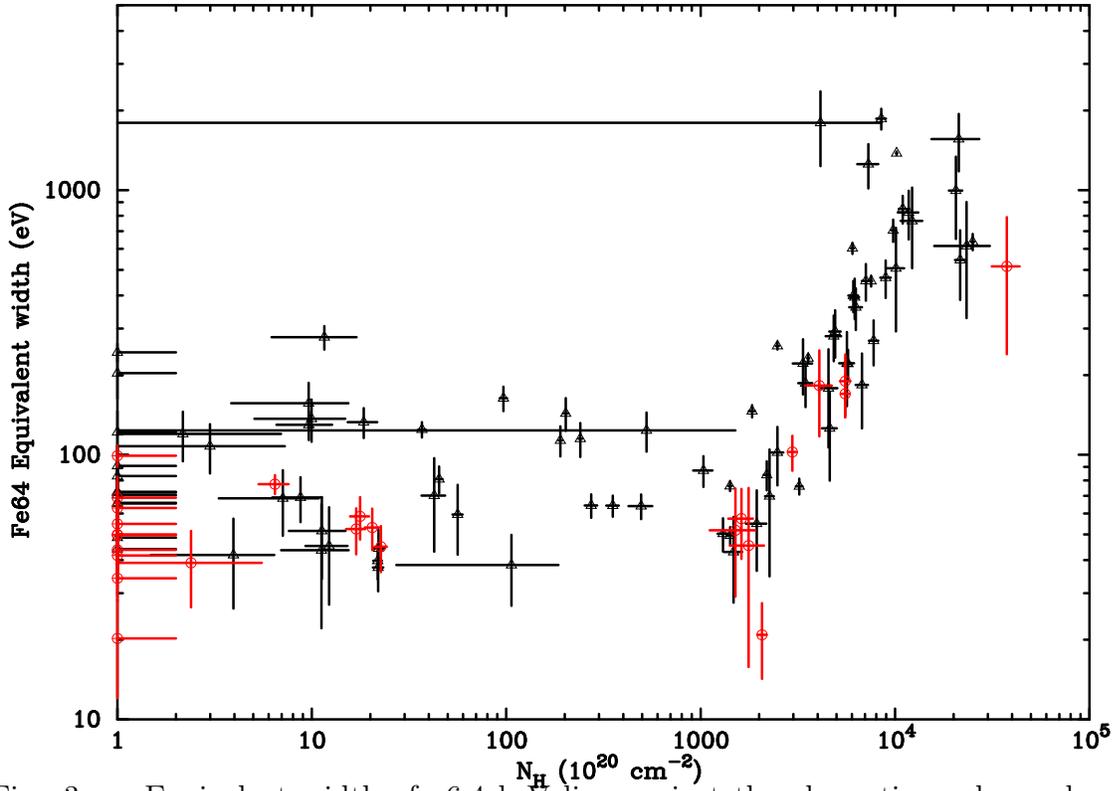}%nh64ew.eps
\caption{ Equivalent width of a6.4 keV line against the absorption column
density. Triangles (black) or circles (red) are AGNs with the
luminosity of $<10^{44}$
and $>10^{44}$ erg s$^{-1}$, respectively. 
\label{nh64ew}}
\end{figure}

Figure \ref{nh64ew} plots an equivalent width (EW) of the 6.4 keV 
line against the absorption column density.
Thanks to effective absorption column density constraints
by the HXD data, the
clearest ever correlation was obtained (e.g. Guainazzi et al. 2005),  
between the EW and
absorption; a positive correlation above $N_{\rm H}=10^{23}$ cm$^{-2}$,
and an almost constant EW around 50--120 eV 
below $N_{\rm H}=10^{23}$ cm$^{-2}$.
The EW distribution for Compton-thin objects are in good agreement with
that of XMM-Newton (Bianchi et al. 2009).
The positive correlation is also in good agreement with the prediction of
the fluorescence line by the Compton-thick torus with various column
density (e.g. Awaki et al. 1991; Ikeda et al. 2009).
Note that a clear correlation within the range $N_{\rm
H}=(0.4-1)\times10^{24}$ cm$^{-2}$ is seen and  
the EW is 400--700 eV at $N_{\rm H}=10^{24}$ cm$^{-2}$.
Since a typical torus structure with
$N_{\rm H}=10^{24}$ cm$^{-2}$ and 1 solar Fe abundance causes 
EW of 600--700 eV for the 6.4 keV line (Awaki et al. 1991; Ghisellini et
al. 1994; Ikeda et al. 2009), 
the Fe abundance of the reflector or absorber is considered to be 
0.7--1.3 solar.

\begin{figure}
\plotone{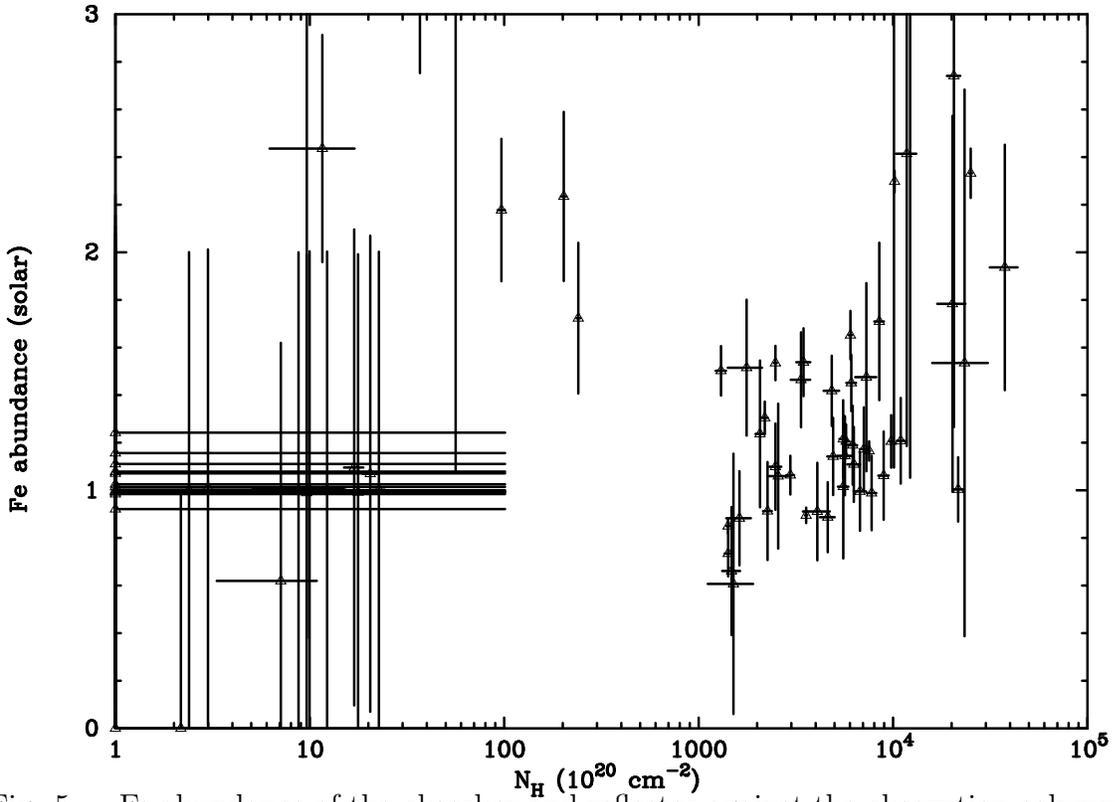}%nhabun.eps
\caption{Fe abundance of the absorber and reflector against the
 absorption column density, derived by fitting with model A.
\label{nhabun}}
\end{figure}

When estimating the $N_{\rm H}$ by fitting the spectra with the model C, 
we fixed the Fe abundance to 1 solar.
However, if we set the Fe abundance to 2 or 0.5 solar, the $N_{\rm H}$ 
systematically changes by a factor of up to 0.83 or 1.03, respectively.
Considering that the relation between $N_{\rm H}$ and EW is almost linear
around $N_{\rm H}=(0.4-1)\times10^{24}$ cm$^{-2}$ (Ghisellini et
al. 1994; Ikeda et al. 2009),
the abundance ratio of Fe to other heavy metals, 
such as O, Ne, Si, and S, is indicated to be close to 0.7--1.3 solar.
Figure \ref{nhabun} shows an Fe abundance of the absorber and reflector
obtained by fitting the spectra with the model A.
In this case, the Fe abundance is determined by the relative strength of
the Fe-K edge and the reflection hump around 20 keV, as demonstrated by
Reeves et al. (2007).
The Fe abundance is well determined to be 0.8--1.5 solar for mildly
Compton-thick objects with the average and variance 1.18 and  0.21
solar, respectively, for objects with $10^{23}<N_{\rm H}<10^{24}$ cm$^{-2}$.
This value is consistent with the estimation from the relation between the
Fe-K line EW and $N_{\rm H}$.

However, several objects deviate from this clear
relation. 
Swift J0601,9-8636, NGC 2273, and NGC 4968 show a very large 
EW of $>1$ keV for $N_{\rm H}<9\times10^{23}$ cm$^{-2}$.
Swift J0601,9-8636 is that reported in Ueda et al. (2007), who
suggested that this AGN is more buried in the geometrically thick torus than
is typical, whereupon a large solid angle of the torus would enhance
the Fe line.
Awaki et al. (2009) reported that the Suzaku spectrum of NGC 2273
contains a significant fraction of the reflection component around the
6.4 line and their sophisticated spectral
modeling reveals a higher $N_{\rm H}$ of $(1.1-1.7)\times10^{24}$
cm$^{-2}$.
meaning that the $N_{\rm H}$ obtained by simple modeling 
might be underestimated.
This situation also applies for NGC 4968, the spectrum of which 
resembles that of NGC 2273.
Some Seyfert 1 AGNs with $N_{\rm H}<10^{21}$ cm$^{-2}$ show a 
relatively large EW of $\sim200$ eV, but this is due to the time
variability of the continuum level; 
good examples of which are NGC 3227 and NGC 5548 which were
observed with Suzaku several times and with the largest EW in the
faintest continuum level.
Note that no objects exist for which the 6.4 keV line intensity varied
among several Suzaku observations (table 1).

\begin{figure}
\plotone{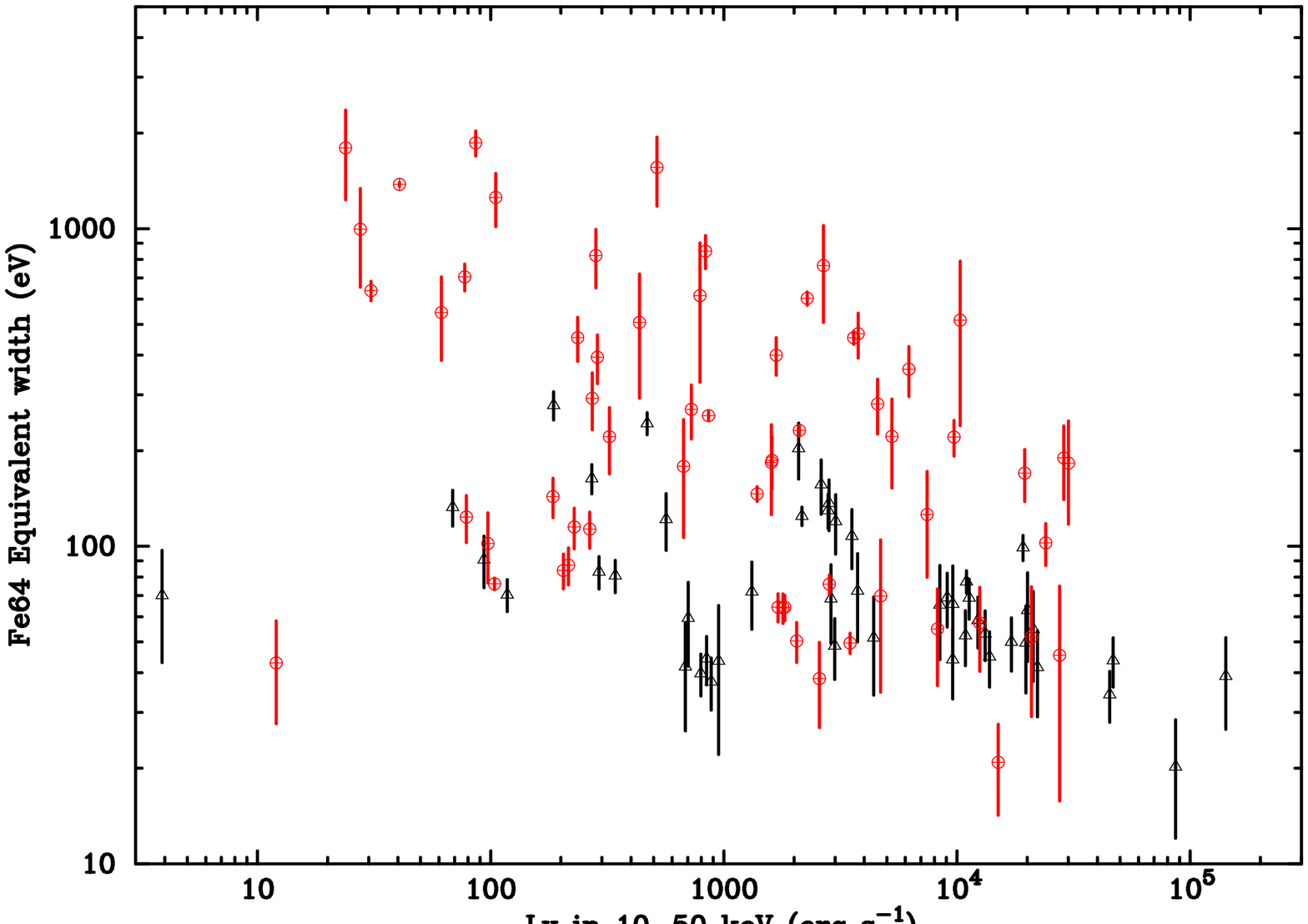}%fe64.eps
\caption{ Equivalent width of the 6.4 keV line against the X-ray luminosity
 (10--50 keV). 
Triangles (black) or circles (red) 
are AGNs with absorption column density of $<10^{22}$ 
and $>10^{22}$ cm$^{-2}$, respectively. 
\label{fe64}}
\end{figure}

On the other hand, certain objects exist with a negligible EW of the 6.4
keV line.
The upper limits of EW in table 1 are obtained by 
fixing the center energy and width
to 6.4 keV and 30 eV, respectively.
High X-ray luminosity objects exhibit a smaller EW of mostly $<80$ eV, 
and some in particular as low as $<25$ eV (IGR J21247+5058,
1H 0419-577, 3C 273, ARK 564, PDS 456, TON\_S180).
This trend of the Baldwin effect (Iwasawa and Taniguchi 1993) 
is clearly seen in figure \ref{fe64}, and Suzaku also 
found a clear trend for the first time even for Seyfert 2 galaxies by
accurately measuring the luminosity
of the direct component above 10 keV.
Note that the sample used contains broad line radio galaxies (3C 120,
3C 390.3, 3C 445, 3C 382, 3C 33, 3C 452, 3C 104) with an X-ray luminosity of
$>10^{44}$ erg s$^{-1}$ (10--50 keV).
Apart from the high luminosity objects, low luminosity objects with 
luminosity of $<10^{41}$ erg s$^{-1}$ also show a low EW, such as M 81
and M 106 (Yamada et al. 2009).
These objects with a high or low luminosity 
are considered to have a reflector with a smaller solid
angle from the central engine by a factor of $>2$ than typical 
Seyfert galaxies.
Except for the above specific situations, there are no extraordinary
objects, deviating from the clear EW-$N_{\rm H}$ relation.

\begin{figure}
\plotone{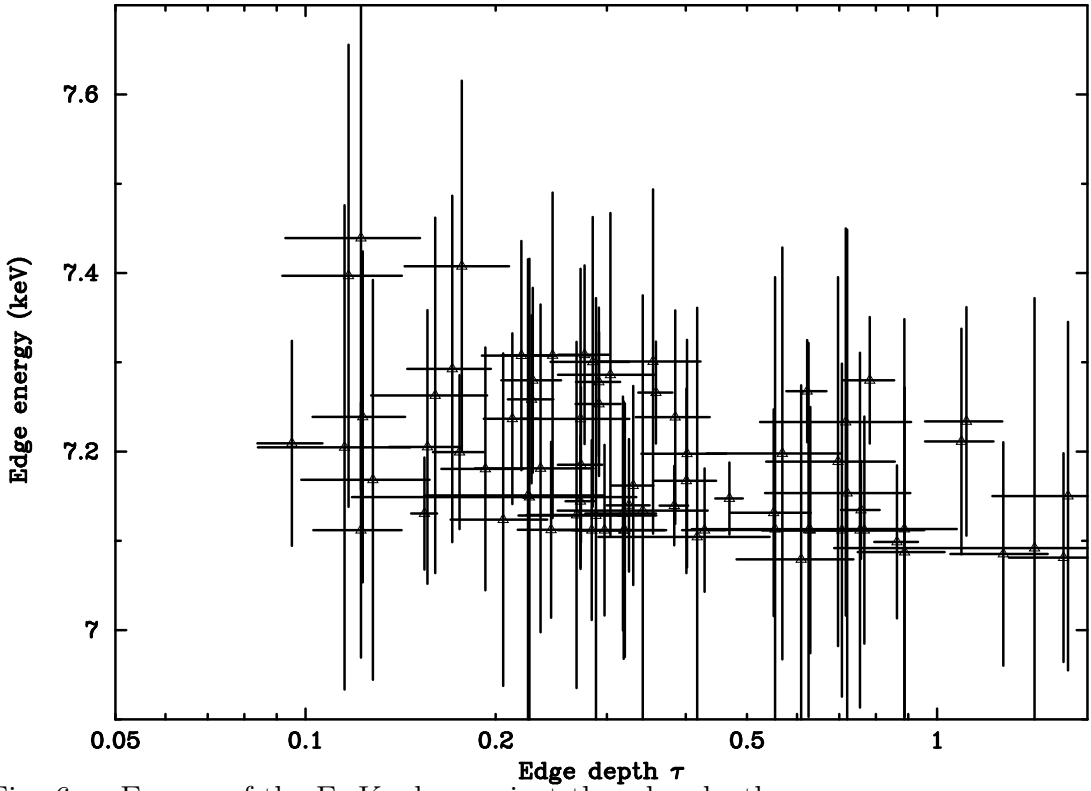}%edge.eps
\caption{Energy of the Fe-K edge against the edge depth.
\label{edge}}
\end{figure}

An Fe-K edge feature around 7.15 keV reveals additional information about
the neutral iron, together with the 6.4 keV line.
The edge energy is more sensitive to the ionization state than the
fluorescence line energy (Makishima 1986; Kallman et al. 2004), 
and the observed
energy is plotted against the edge depth $\tau$ in figure \ref{edge} 
for objects with small error bars, 
which reveals observed edge energy of around 7.2$\pm$0.1 keV and 
an indicated ionization state of $<9$.
Limited to $\tau>0.4$, the edge energy is determined to be below 7.2
keV, tightly constraining 
the ionization state to $<6$ for Compton thicker objects.
Note that there is a hint of a higher edge energy for lower $\tau$ 
objects. 
The average edge energy is 7.204$\pm$0.022 and 7.110$\pm$0.033 keV
for objects with $\tau<0.3$ and $>0.4$, respectively.
The energy of the Compton-thicker objects is consistent with that for the
neutral Fe-K edge of 7.112 keV (X-ray Transition Energy Database), while
the energy of Seyfert 1 galaxies indicates the weak 
ionization state of 3--7 or the contribution of the disk/absorption line.

\begin{figure}
\plotone{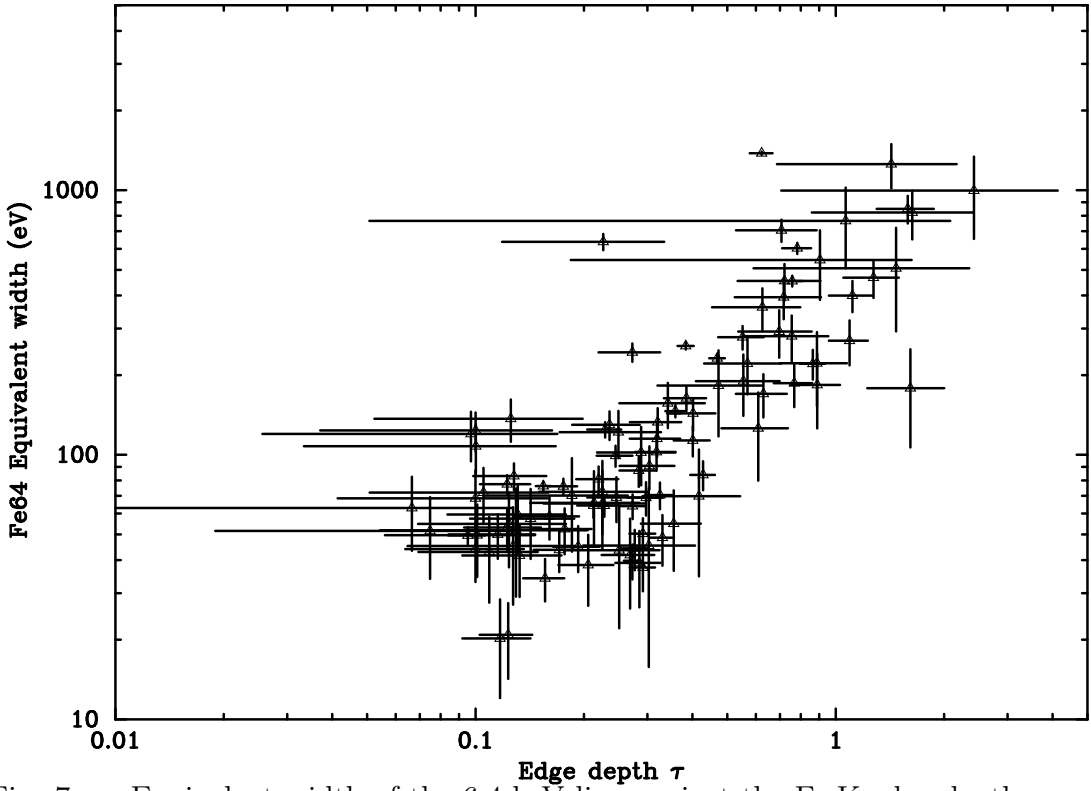}%tau64ew.eps
\caption{ Equivalent width of the 6.4 keV line against the Fe-K edge depth.
\label{tau64ew}}
\end{figure}

Line EW and edge depth $\tau$ should suggest a linear 
correlation if the
absorber and reflector have the same column density, ionization
degree, and iron abundance, following
$\tau\sim\frac{N_{\rm}}{4.2\times10^{23} {\rm cm}^{-2}}$.
Figure \ref{tau64ew} shows the plot between an edge depth and a 6.4 keV line EW.
Both values correlate relatively well, indicating identical origin,
while some outliers exist at an edge depth of 0.2--0.5 and a high EW of
$>400$ eV.
These objects are almost Compton-thick (such as NGC 4945), 
the spectrum of which is dominated by the reflection component.
Therefore, the deviation is due to the fact that for these objects,
the edge is mainly caused by reflection, 
whereas for other Compton-thinner objects, 
it is mainly caused by absorption.
The $\tau$ scatter for low EW objects is due to the origin of the 
absorption differing from that of the reflection as described in the 
discussion.
Another possibility is being
affected by a disk line or weak unresolved lines around their shallow 
edge structure.

\subsection{Ionized Fe-K Emission and Absorption Lines}

\begin{figure}
\plotone{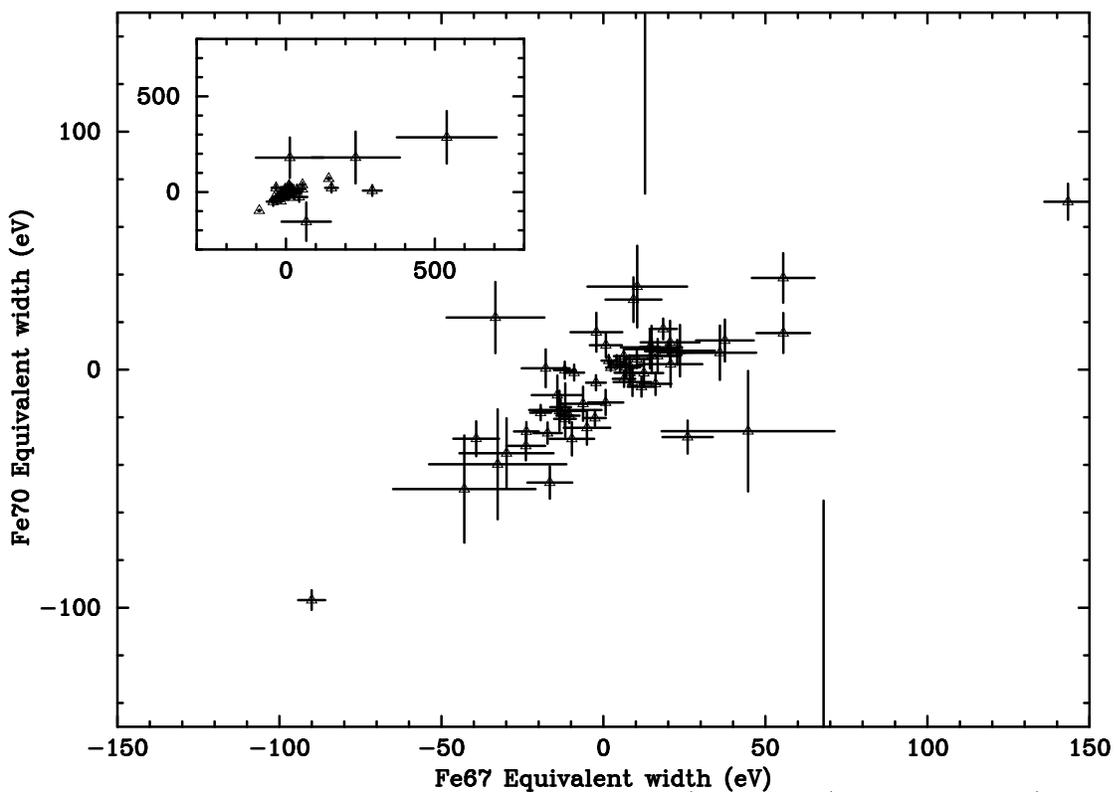}%fe6770.eps
\caption{ Plot of the equivalent width of 6.7 keV (horizontal) and 7.0 keV
 (vertical) lines. The negative values represent the absorption line. The
 inset is a plot where the horizontal and vertical range is enlarged.
\label{fe6770}}
\end{figure}

Thanks to the unprecedented signal-to-noise ratio of the Suzaku data 
for many AGNs, it is revealed that ionized emission and absorption 
lines with equivalent widths
of 5--60 and 5--40 eV, respectively, are very common
for Seyfert galaxies, regardless of whether type 1 or 2 objects.
This is in good agreement with the results for Compton-thin objects, 
based on the XMM-Newton data (Bianchi et al. 2009).
Figure \ref{fe6770} shows the relation of EWs between 6.7 and 7.0 keV 
lines.
Here, although the line intensities correlate well with each other, 
some deviating objects do exist --
some objects showing either a 6.7 keV emission line or a 7.0 keV
absorption line (MRK 766, NGC 7582).
SWIFT J1628.1+514 shows a 6.7 keV absorption line and a 7.0 keV emission line.
The strongest emission lines are seen for Compton-thick Seyfert 2
galaxies (NGC 1068, NGC 4945, MRK 273, CIRCINUS GALAXY, and IRAS 19254-7245).
For NGC 1068 and NGC 4945, the 6.7 keV line is strong but the 7.0 keV
line less so.
Conversely, relatively strong absorption lines are seen for NGC 1365,
whose EWs of 6.7 and 7.0 keV absorption lines are around 100 eV.
In contrast with the 6.4 keV line, 
NGC 3227, NGC 5548, and MRK 766 show a variability of ionized Fe-K lines, 
in such a way that the absorption line at 6.7
keV appeared with increasing flux over a time scale
of $\sim10$ days. 
Such variability also emerged for NGC 3783 by XMM-Newton 
(Reeves et al. 2004).

\begin{figure}
\plottwo{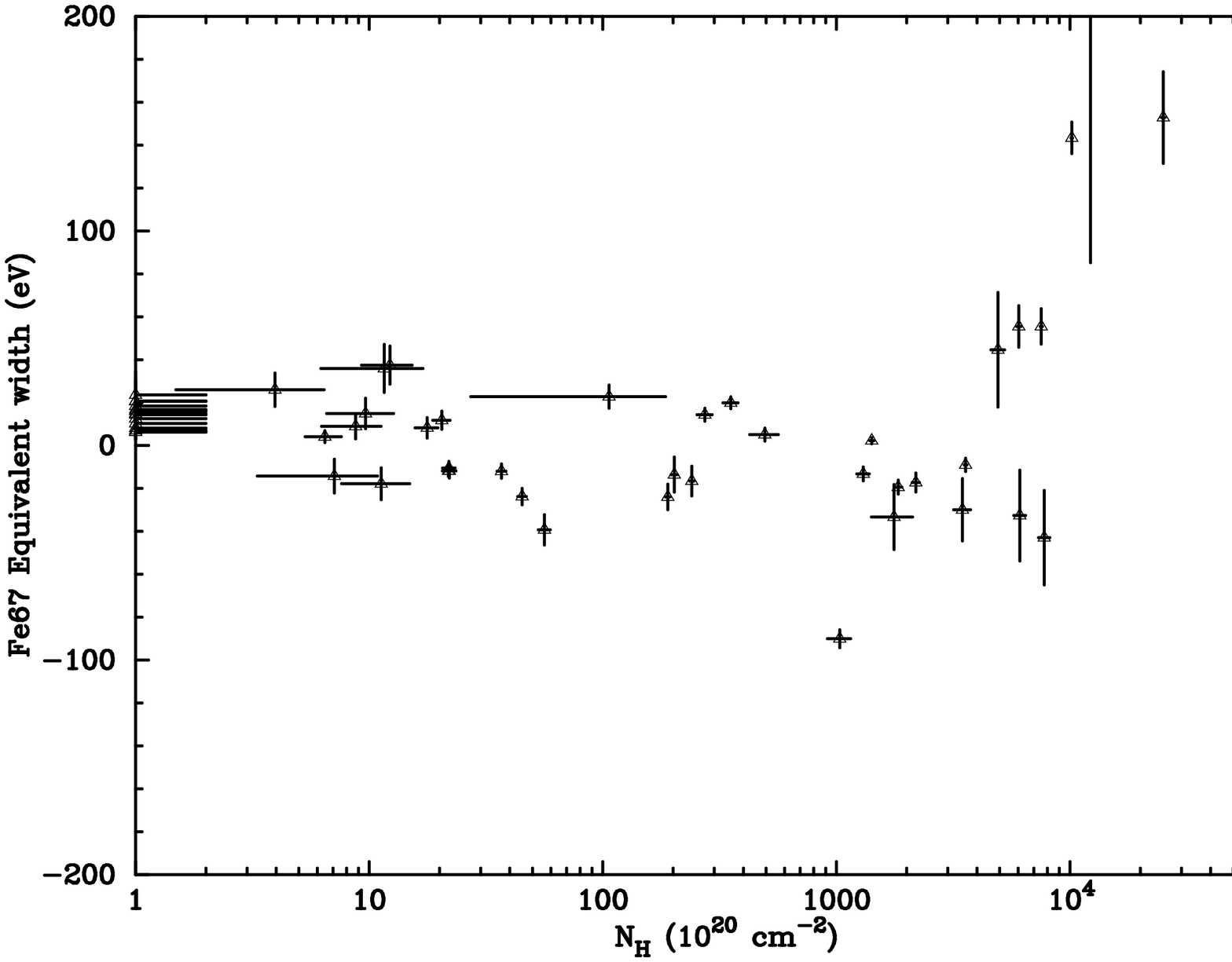}{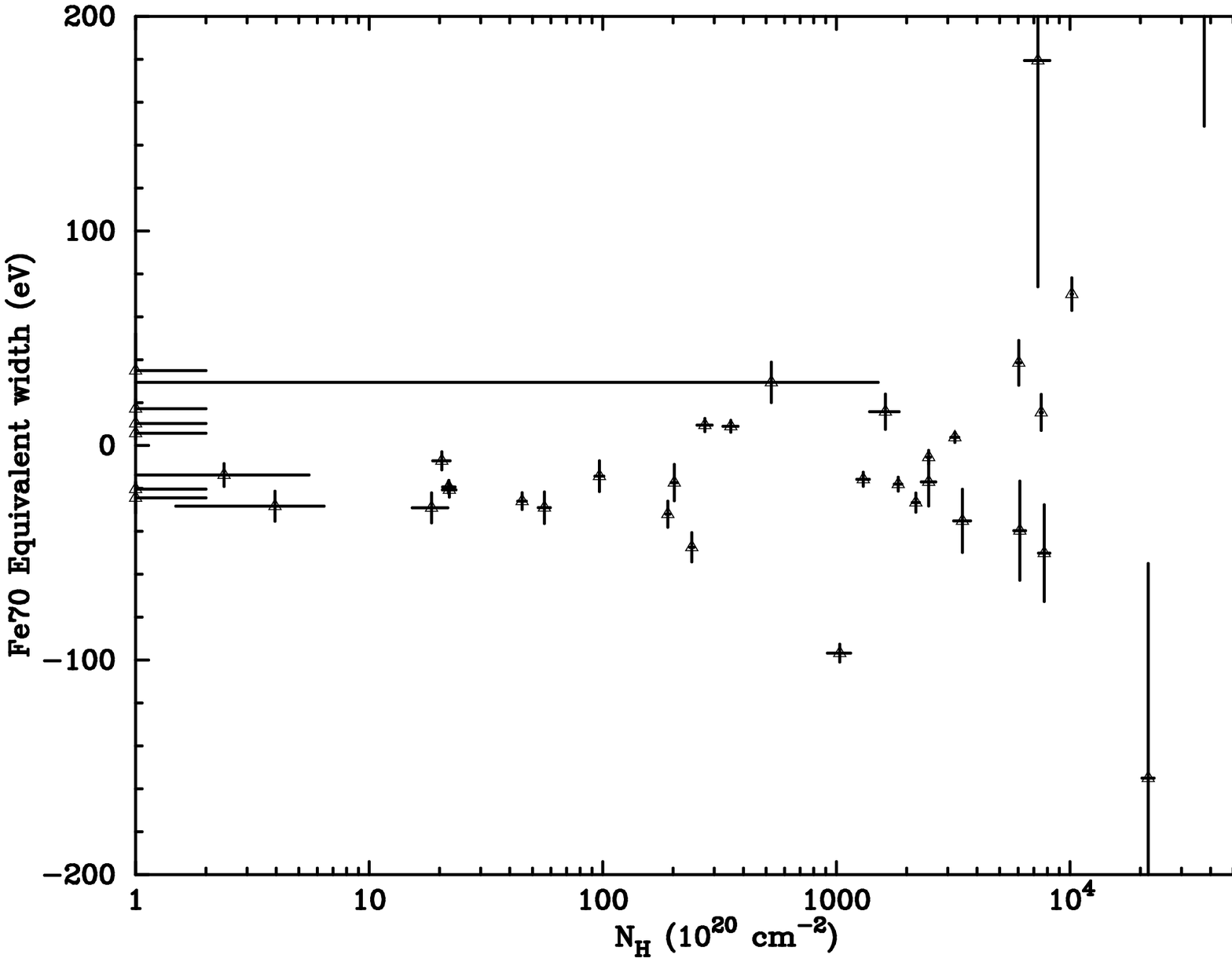}%{nh67ew.eps}{nh70ew.eps}
\caption{ Equivalent width of 6.7 keV (left) and 7.0 keV (right) 
line against the absorption column density of the cold material.
\label{fe6770nh}}
\end{figure}

Figure \ref{fe6770nh} shows a plot of an equivalent width 
of 6.7 and 7.0 keV lines against an absorption column density of cold
material.
Seyfert 1 galaxies show a smaller EW of $<20$ eV for both emission and
absorption lines, while
Compton-thick objects show a large EW for the emission lines; 
sometimes $>100$ eV, with
the trend for the emission line resembling that of the 6.4 keV line;
the EW is almost constant below $N_{\rm H}=10^{23}$ cm$^{-2}$ but increases
rapidly above $N_{\rm H}=10^{23}$ cm$^{-2}$, indicating that
the geometrical distribution of highly ionized material is similar to that
of the neutral one but the covering solid angle is around 10\% of the 
cold material.
This trend is not seen for the absorption line, but a relatively large
EW tends to be observed for objects with larger $N_{\rm H}$.

\begin{figure}[hpbt]
\plottwo{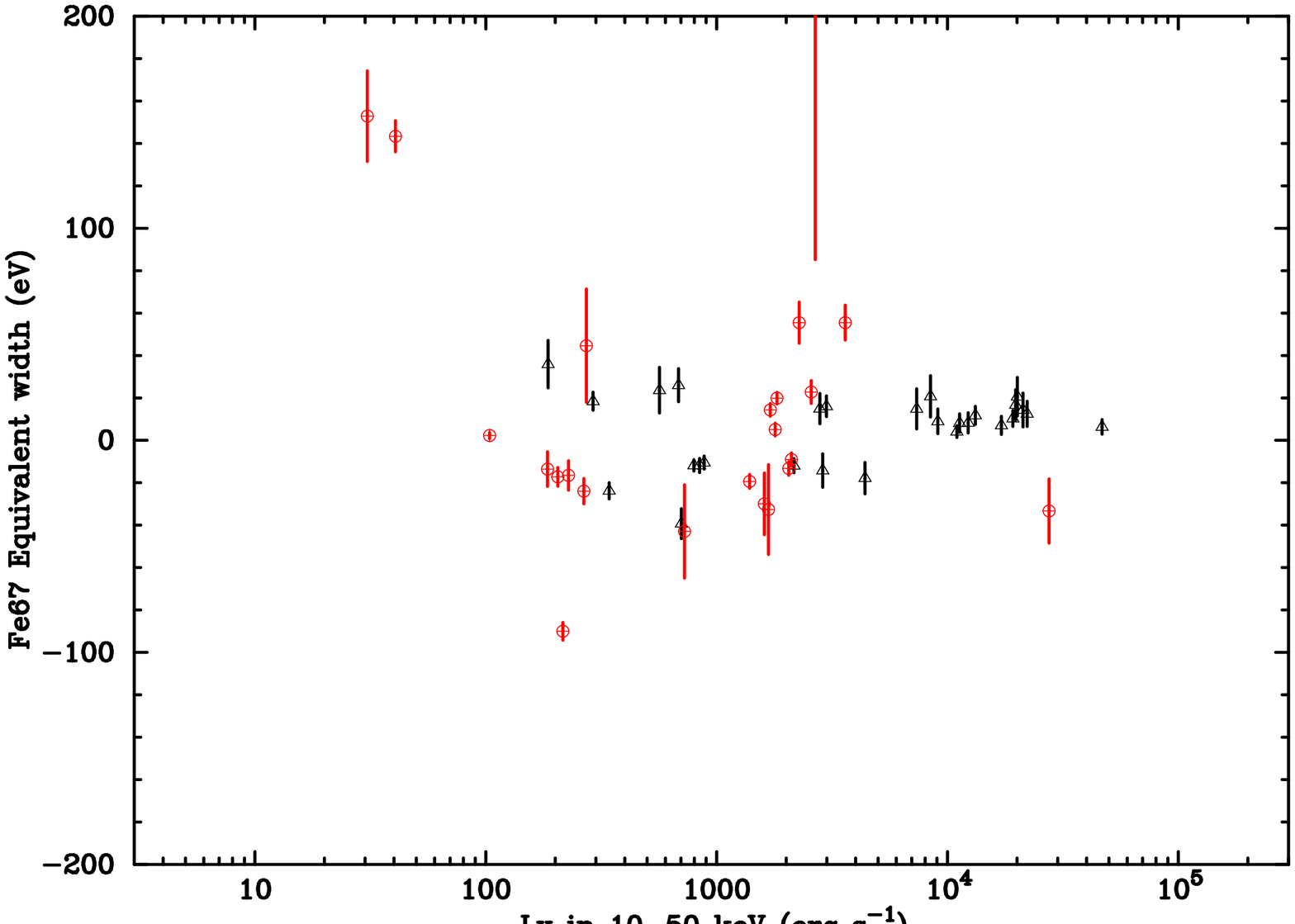}{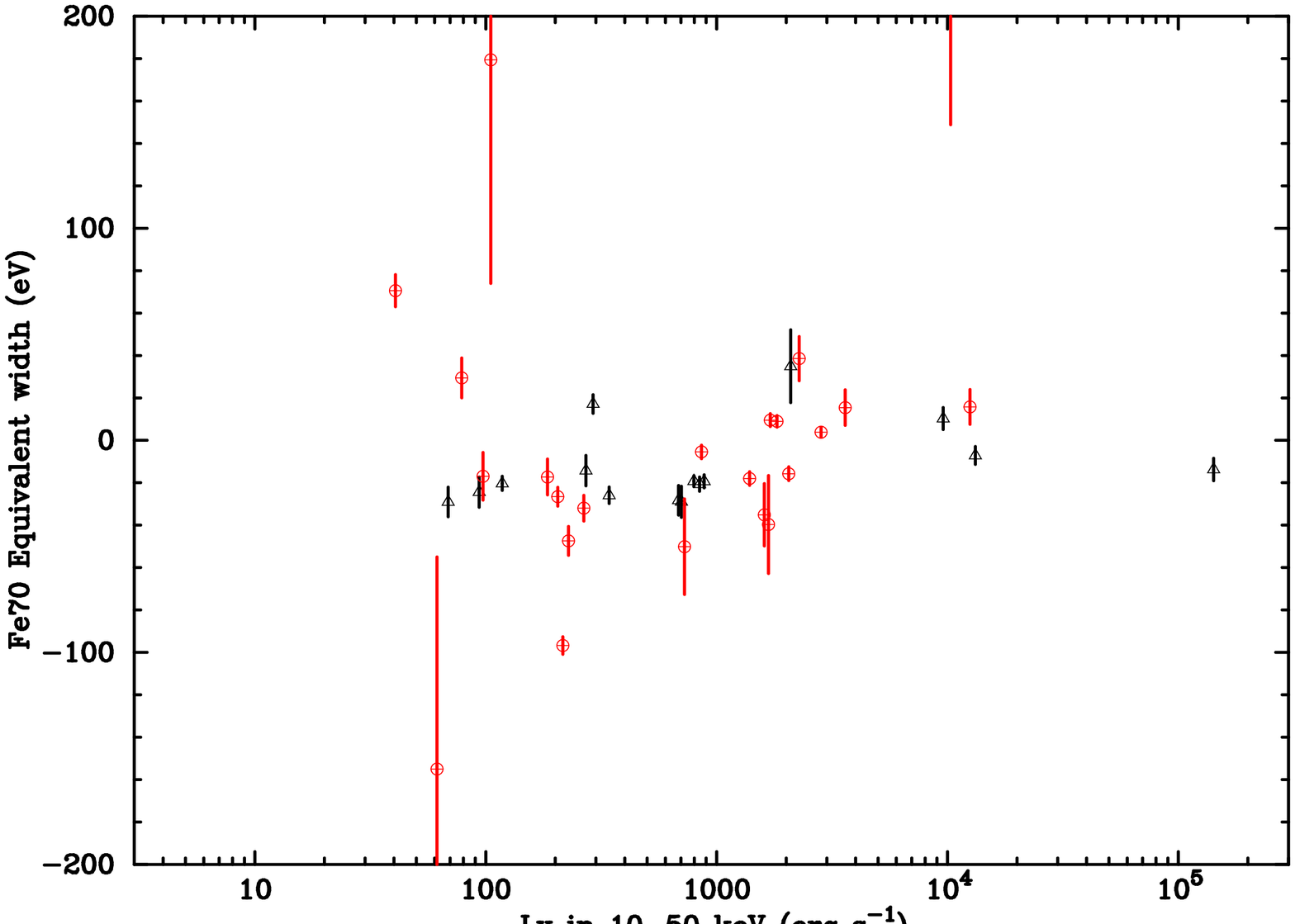}%{fe67.eps}{fe70.eps}
\caption{ Equivalent width of 6.7 keV (left) and 7.0 keV (right) 
line against the X-ray luminosity (10--50 keV). 
Triangles (black) or circles (red) 
are AGNs with 
absorption column density of $<10^{22}$ 
and $>10^{22}$ cm$^{-2}$, respectively. 
\label{fe6770lx}}
\end{figure}

Figure \ref{fe6770lx} shows a plot of an equivalent width of 
6.7 and 7.0 keV lines against an X-ray luminosity (10--50 keV).
There is a trend of smaller equivalent width for higher X-ray
luminosity.
An average of the absolute value of EW, which is taken for objects 
whose line is significant at
1.5$\sigma$ level as shown in table 1, is 22 and 11 eV below and
above $L_X=5\times10^{43}$ erg s$^{-1}$, respectively, for the 6.7 keV 
line.
The average for the 7.0 keV line is 24 and 10 eV below and
above $L_X=5\times10^{43}$ erg s$^{-1}$, respectively.
Correlation coefficients are calculated at -0.37 
and -0.43 for the 6.7 and 7.0 keV lines, respectively, in comparison with
-0.39 for the 6.4 keV line. Therefore, the
coefficients are almost comparable to that of the 6.4 keV line which
shows a clear Baldwin effect, despite the significant errors of the 6.7 
and 7.0 keV lines.
This is the first hint of the Baldwin effect for ionized Fe-K lines.
These correlations suggest that the physical properties of the ionized material
are related with the neutral material and the nuclear luminosity.

\section{Discussion}

\subsection{Properties of the Neutral Material Emitting 6.4 keV line}

Tne center energy of the 6.4 keV line and the Fe-K edge
strongly constrain the ionization states of the absorber/reflector to less
than 12 and 6, respectively.
This is the most severe constraint to date, thanks to the Suzaku capability.
The latter corresponds to the ionization parameter of
$\xi=\frac{L}{nR^2}\leq0.1$, where $L$ is the luminosity of the central
engine, $n$ the matter density, and $R$ the distance to the matter
(Kallman et al. 2004).
The width of the 6.4 keV line is constrained to $<3300$ km s$^{-1}$ 
or $<$70 eV for objects with well-determined values, 
suggesting that the origin of
the 6.4 keV line is a broad-line region or a torus.
The line equivalent width and the edge depth correlate well with each
other, indicating that both features are due to the same origin.
Therefore, 
considering the above constraint of $\xi$ and the 
column density $N_{\rm H}<nR$,
the 6.4 keV line likely comes from the neutral
matter at $>3L_{43}N_{\rm H,25}^{-1}$ pc away from the nucleus, where $L_{43}$
and $N_{\rm H,25}$ represent the luminosity and column density in units of
$10^{43}$ erg s$^{-1}$ and $10^{25}$ cm$^{-2}$, respectively.
The reduced time variability of the 6.4 keV line is also 
consistent with this picture.
Since the inner torus radius is reported at
0.03$L_{43}^{0.5}$ pc (Suganuma et al. 2006), 
the reflection matter is not the outer disk but the torus.
The absorption column density reportedly varied rapidly within
one day (e.g. Elvis et al .2004) or several weeks (Risaliti
et al. 2005).
Although the blob-like structures of the Compton-thick
absorber could cause such short-term variability, the phenomena could
also be attributable to the 
underlying reflection component, even
if only the direct nuclear component varies.
Therefore, the fast time variability of the absorption column density 
must be focused on by carefully considering the reflection
component.

Seyfert 1 galaxies and Compton-thin Seyfert 2 galaxies with
$N_{\rm}<10^{23}$ cm$^{-2}$ show a similar EW of 50--120 eV for the 6.4
keV line, regardless of the absorption column density.
By simple estimation,
a reflection component has an EW of $\sim$1000 eV for 1 solar abundance,
$N_{\rm}<5\times10^{23}$ cm$^{-2}$,
and the input powerlaw photon index of 2.
Considering the transmission and reflection efficiency of the reflector
with a solid angle of $\Omega$ and a column density of $N_{\rm H}$, 
the sum of the direct and reflection components reveals
an Fe-K EW of $\sim300\frac{\Omega}{4\pi}\frac{N_{\rm H}}{4\times10^{23}
{\rm cm}^{-2}}$ eV.
Therefore, a Compton-thin absorber cannot create the observed large EW of
50--120 eV, 
and the Compton-thick reflector with $N_{\rm H}>10^{23}$ cm$^{-2}$
should exist around the nucleus at the off line of sight even for Seyfert 1 
galaxies, with possible candidates including 
accretion disk or dust torus
(Awaki et al. 1991; Guainazzi et al. 2005).
Since the constraint of the ionization state of reflectors requires the
location of $>3L_{43}N_{\rm
H,25}^{-1}$ pc from the nucleus, the reflector for
Seyfert 1 galaxies is not the disk but the torus.
The EW of 50--120 eV indicates a solid angle of $(0.2-0.7)\times4\pi$
for the Compton-thick torus with $N_{\rm H}>4\times10^{23}$ cm$^{-2}$
(Ghisellini et al. 1994; Ikeda et al. 2009).
As discussed in Ghisellini et al. (1994), the reflection by the inner disk
also contributes to the EW by 100--150 eV (Matt et al. 1992).
However, various ionization states, Doppler, and relativistic effects
would smear the Fe-K line features, meaning that 
the Fe-K line from the inner disk is not single nor narrow.

A clear correlation between the column density and the 6.4 keV line EW
indicates that the torus is an absorber with a large solid angle and 
that the torus column density varies within the range
$10^{23-24.5}$ cm$^{-2}$ on an object by object basis.
Large EWs of 6.4 keV at $N_{\rm H}<10^{23}$ cm$^{-2}$
mean that the reflection material differs from the absorber.
Therefore, the absorber below 10$^{23}$ cm$^{-2}$ is not a torus, 
but a low column material toward the off-torus direction.
In the case of $N_{\rm H}=10^{22}$ cm$^{-2}$, the low ionization state of
$\xi<1$ constrains the distance of the absorber to $>0.3$ kpc for 
luminosity of $10^{43}$ erg s$^{-1}$.
Therefore, the matter is likely to be part of the interstellar medium 
(Matt 2000; Guainazzi et al. 2005).

\begin{figure}[hpbt]
\plotone{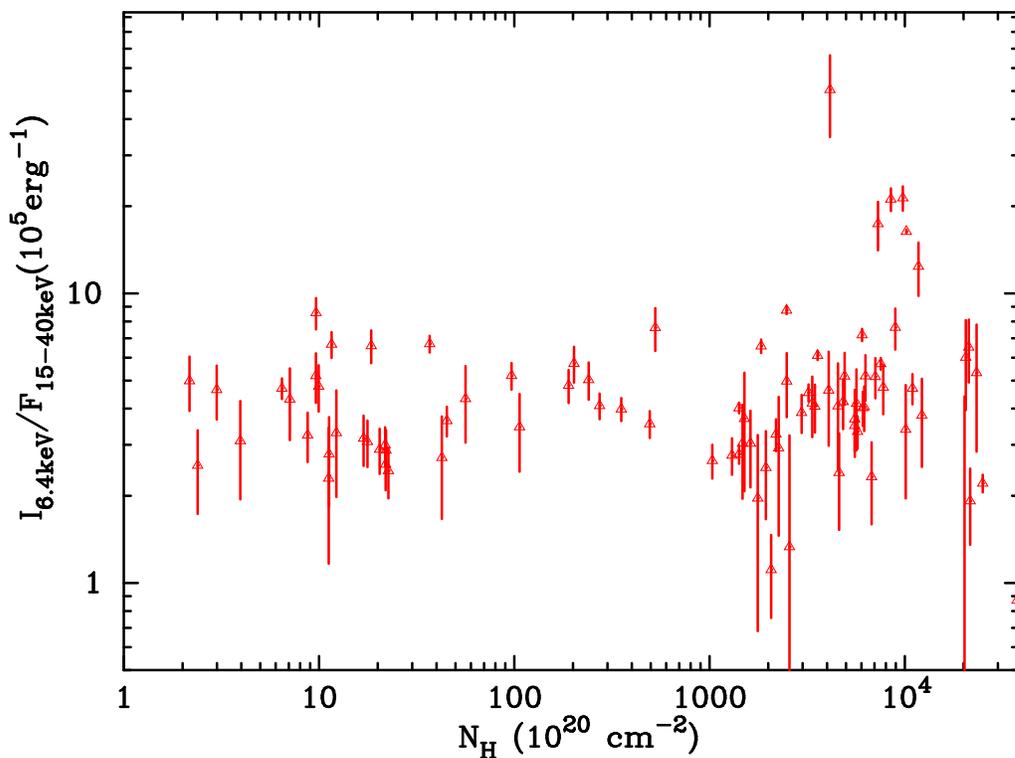}%{nh64pofl.eps}
\caption{ The 6.4keV line intensity ratio against the flux in 10--50
 keV, plotted against the absorption column density. 
\label{nh64pofl}}
\end{figure}

Based on the broadband X-ray spectroscopy, it has been reported that 
the relative reflection component for Seyfert 2 galaxies 
is larger than that of Seyfert 1 galaxies (Malizia et al. 2003).
However, there is no trend of higher EW for larger column density below
$10^{23}$ cm$^{-2}$.
Above $10^{23}$ cm$^{-2}$, the correlation between the EW and column
density constrains the solid angle to $(0.2-0.5)\times4\pi$, comparing
the relation obtained by the simulation (Ghisellini et al. 1994; 
Ikeda et al. 2009).
This is not different from $(0.2-0.7)\times4\pi$ of Seyfert 1 galaxies.
Figure \ref{nh64pofl} shows the 6.4 keV line intensity ratio against
the flux in 10--50 keV,
which corresponds to the emission efficiency of the 6.4 keV line
against the direct nuclear emission, since
the observed flux in 10--50 keV is almost equivalent to the flux of the direct
nuclear emission, as described in \S3.
Therefore, this ratio is not significantly affected by the 
absorption like the equivalent width and is
calculated as $(4.3-7.9)\times10^5$ erg$^{-1}$ 
for the photon index of
1.5--2.2, a solid angle of $\Omega_{0.2}=0.2\times4\pi$, a reflector column
density of $1\times10^{24}$ cm$^{-2}$, Fe abundance of 
$A_{\rm Fe,1solar}=$1 solar, Fe-K
fluorescence yield of 0.34, and a K$\alpha$ fraction of 0.88.
The observed ratio is almost within the range $(2.5-7.0)\times10^5$ 
erg$^{-1}$ for various absorption column densities, suggesting that
the 6.4 keV emitter has a geometrical structure with
$A_{\rm Fe,1solar}\Omega_{0.2}=0.5-1.0$ commonly for many Seyfert
galaxies, and that no clear difference exists between Seyfert 1 and 2
galaxies.
Some Compton-thick objects exhibit a large ratio above $10\times10^5$ 
erg$^{-1}$; 
Circinus Galaxy, NGC 1068, NGC 2273, NGC 4968, ESO 323-G032, and Swift
J0601.9-8636, indicating that no direct nuclear emission is observed
even above 10 keV, namely, that they are reflection-dominated.
Another possibility is that the solid angle of torus is exceptionally
large as suggested for Swift J0601.9-8636 (Ueda et al. 2007).
The ratio of NGC 4945 is $(2.20\pm0.15)\times10^5$ erg$^{-1}$ for 
$N_{\rm H}=2.5\times10^{24}$ cm $^{-2}$.
When correcting the absorption, the flux becomes 1.5 times higher, and
the ratio is reduced to $1.46\times10^5$ erg$^{-1}$, 
which is consistent with the suggestion that NGC 4945 has a small solid
angle of torus (Itoh et al. 2008).
This plot is therefore very useful to consider the torus solid angle
and distinguish whether or not the continuum is reflection-dominated.

The relation between the absorption column density and the EW of the 6.4
keV line, together with the ratio of the Fe-K edge depth and the 
reflection hump around 20 keV, 
indicates that the Fe abundance of the reflector and absorber 
is around 0.7--1.3 solar.
Detailed spectral modeling of the Suzaku spectrum of the Compton-thick 
MRK 3, based on the simulation, is almost consistent with 1 solar
abundance (Ikeda et al. 2009).
For the Compton-thick NGC 1068, 
the Fe abundance was reported to be 2--4 solar, based on detailed 
analysis of
the neutral and ionized Fe-K lines (Matt et al. 2004; Bianchi et al. 2005).
An enhanced starburst activity of NGC 1068 would cause an overabundance
(Bruhweiler et al. 1991), but this is an open issue.

\subsection{Properties of Ionized Material}

In most cases, the EW of the ionized Fe-K emission lines is 5--50 eV.
Referring to the K$\alpha$ fluorescence yield $\sim$0.7 
of Fe$^{+25}$ (Krolik \& Kallman 1987),
the EW of the emission line due to fluorescence of the
ionized material, which is not so Compton-thick, 
becomes $300\frac{\Omega}{4\pi}\tau_e$ eV, where
$\tau_{\rm e}$ is the photoelectric optical depth of the ionized matter.
Resonance scattering by the material outside the line of sight could
become the dominant source of the emission line 
in the optical-thin regime ($\tau_{\rm e}<<$1) 
(Matt et al. 1994).
Therefore, both fluorescence and resonance scattering contribute to
the Fe-K emission lines, but the fluorescence dominates for higher
column density.
Bianchi and Matt (2002) calculated the EW of ionized Fe-K lines 
from both fluorescence and resonance scattering, and their results
indicate that the observed EW of 5--50 eV requires a column density of
$>10^{21-23}$ cm$^{-2}$.
In their geometry, the solid angle of the ionized material is
$0.26\times4\pi$.
About 30\% of objects exhibit an ionized emission line at 1.5$\sigma$
level (90\% confidence level),
indicating that the ionized material of the above column density
subtends about 30\% of $4\pi$, which resembles the situation in 
Bianchi and Matt (2002).

On the other hand, the EW of ionized Fe-K absorption lines is often 5-40
eV.
Chandra HETG observations of some Seyfert galaxies reported the
width of Fe-K absorption lines to be around several 1000 km s$^{-1}$ 
(Kaspi et al. 2002; Risaliti et al. 2005).
According to the growth curve of the Fe-K absorption lines (Bianchi et
al. 2005), 
the column density $N_{\rm H}$ of the absorption matter
toward the line of sight is $10^{21.5-22.5}$ cm$^{-2}$, corresponding to
$\tau_{\rm e}=0.01-0.1$.
which suggests that the emission and absorption lines are both due to 
the same matter.
No Compton-thick objects exhibit absorption lines, which is attributable
to the significant reduction in the direct nuclear emission
transmitting the ionized material around the Fe-K
lines and the domination of the reprocessed component.

The ionization parameter $\xi=\frac{L}{nR^2}$ is considered to be $10^{3-4}$.
Therefore, taking $L=10^{42-44}$ erg s$^{-1}$, the distance $R$ 
from the nucleus to the ionized matter is
constrained to $R<10^{16-19}$ cm.
This is consistent with a time scale showing variability of the Fe-K 
absorption lines of around several weeks for NGC 5548 and NGC 3227.
Assuming $N_{\rm H}=10^{22}$ cm$^{-2}$, the density is $n=10^{3-6}$
cm$^{-3}$ and
while the highly ionized matter is located at the broad line region,
its density is much lower and the ionization state much higher than the
broad-line-emitting cloud.

Some Compton-thick objects, such as NGC 1068, NGC 4945,
Circinus Galaxy, and IRAS 19254-7245, exhibit strong 6.7 and/or 
7.0 keV emission lines with EW$>100$ eV.
In these cases, the ionized material should subtend a large solid angle of 
$\Omega>(0.2)\times4\pi/\tau_{\rm e}$.
If the ionized matter emitting the emission line is the same as that
causing the absorption line, the column density is $N_{\rm
H}=10^{21-22}$ cm$^{-2}$ whereupon the $\Omega$ becomes comparable to or
exceeds $4\pi$ in most cases.
Therefore, the material emitting ionized Fe-K lines for these objects 
differs from that of Compton-thinner objects.
Assuming column density of $N_{\rm H}=4\times10^{23}$
cm$^{-2}$ and an ionization parameter of $\xi=10^3$, the distance from
the nucleus is constrained to be $R<2\times10^{16}L_{43}$ cm.
In the case of NGC 4945 and Circinus Galaxy with $L\sim10^{42}$ erg s$^{-1}$, 
$R$ becomes $<2\times10^{15}$ cm or $<1$ light day or $10^{2-4}$ of the
black hole Schwarzschild radius with a mass of $10^{6-8}$ M$_{\odot}$.
The maser emission for these objects indicates that the inclination of the disk
is nearly edge-on.
Therefore, the outer disk region appears to be somewhat vertically
extended and highly photo-ionized, with nuclear emission
visible through this region.
These parameters resemble those of the ionized matter of 
Fe-K absorption lines for NGC 1365 with 10 times higher luminosity, but
it exhibits a blue shift of 500--1500 km s$^{-1}$.
The rapid increase of EW of the ionized Fe-K emission lines at
$N_{\rm H}>10^{23}$ cm$^{-2}$ indicates that the column density of
the ionized material increases alongside that of the cold material.
This situation is explained by the partial photoionization of the outer
disk region.

\subsection{Luminosity Dependence of the Environmental Material}

High luminosity AGNs are shown to systematically exhibit a smaller
EW for the 6.4 keV line and also possibly for the 6.7 and 7.0 keV lines.
In the sample used, high luminosity AGNs include radio-quiet quasars,
narrow-line Seyfert galaxies, and broad-line radio galaxies.
This trend is seen even when Seyfert 2 galaxies are excluded, which show a
smaller $N_{\rm H}$ for higher luminosity (e.g. Ueda et al. 2003).
These indicate that both of neutral material (torus) and
ionized matter are severely irradiated by the bright nucleus and a
significant portion of the material becomes fully ionized and evaporated.
This trend of the 6.4 keV line 
for Seyfert 2 galaxies matches the deficiency of 
Compton-thick objects for higher luminosity.
Somewhat higher Fe-K edge energy for AGNs with lower absorption column density
suggests that the inner part of torus begins to be photoionized.

Low luminosity AGNs also show a smaller EW of the 6.4 keV line, but
the ionized Fe-K line is not weak.
It is suggested that advection-dominated accretion flow is formed 
when the accretion rate is much smaller than the 
Eddington limit (Narayan \& Yi 1994, Chiaberge et al. 2006).
In this case, the torus which supplies accretion material is
considered smaller, and the low-density accretion flow is
photoionized.

These observational features for ionized Fe-K lines can be compared with
the inner radius of dust torus against the AGN optical
luminosity (Suganuma et al. 2006), where light travel time $\delta t$
from the nucleus to the inner radii of the torus is proportional to the 
square of luminosity as $\propto L^{0.5}$. 
When the luminosity exceeds several $10^{44}$ erg s$^{-1}$, 
the inner region of the dust torus evaporates and 
its inner radius becomes as large as 100 light days.
Also,
highly ionized material, which exists at a smaller radius than the
torus, is fully ionized, whereupon emission or absorption lines
cannot be observed.
This picture matches the tendency of lower
cold absorption column density for high luminosity AGNs
(e.g. Ueda et al. 2003).
Furthermore the higher the luminosity of the radiation, the greater the
propensity for massive outflow
to occur, as observed in some high luminosity AGNs.
The luminosity-dependence of these
properties, regardless the black hole mass scatter by two orders of
magnitude, indicates that the torus is related to the structure of the
parent galaxy rather than the central black hole.

\subsection{Summary}

We systematically analyzed the Fe-K line features of 88 Seyfert galaxies
with high-quality Suzaku data, and derived the properties of matter
around the nucleus.
Based on the constraint of the ionization state and the relation among EW,
column density, and edge depth, 
neutral matter causing a 6.4 keV line and 7.1 keV edge is likely to be
the torus with a solid angle of typically $(0.15-0.4)\times4\pi$ and an
Fe abundance of 0.7--1.3 solar for
both Seyfert 1 and 2 galaxies, but
certain high or low luminosity objects require a very small solid angle of
torus, possibly due to the small amount of accreting material or the
significant photo-ionization of the torus.
No systematic difference emerges between Seyfert 1 and 2 in terms of
properties. 
We propose a new measurement value 
which is useful to consider the above issues; the
6.4 keV line intensity ratio against the 10--50 keV flux.
Highly ionized Fe-K lines are found for 20--30\% of Seyfert galaxies.
Their ionization state and EW are consistent with the picture that
highly ionized matter is located at the broad line region and they exists
with a much lower density than the broad-line-emitter.
A large EW of He-like Fe-K lines for certain Compton-thick objects 
indicates the contribution of the photoionized outer disk as viewed edge-on.

In the near future, high-resolution X-ray spectroscopy by micro
calorimeter will become available, which will allow the nuclear region
to be probed more
precisely by an order of magnitude than by current X-ray CCD observations.
Compton shoulder of the neutral Fe-K line and line width are powerful
parameters to 
constrain the location and geometry of neutral matter.
Precise measurements of the line and edge features, including satellite
lines, reveal information concerning the chemical state of the matter, 
which thus allows us to probe the dust state.
The fine structure of the ionized emission or absorption lines is also 
a strong tool to constrain
the matter density and precisely determine the ionization state, while 
the line profile and center energy are important to derive information of
matter velocity.
Such measurements are currently performed for only a few objects by long-look
Chandra HETEG observations.
Disk line features can also be unambiguously tested,
since their features can be separated from other Fe-K line and edge features.
ASTRO-H SXS is the first to perform such measurements for many AGNs
and furthermore track their time variation.

The authors wish to thank all members of the Suzaku Science Working
Group, for their contributions to the instrument preparation, spacecraft
operation, software development, and in-orbit calibration. This work is
partly supported by Grants-in-Aid for Scientific Research by the
Ministry of Education, Culture, Sports, Science and Technology of Japan 
(20340044).

\clearpage

%\begin{deluxetable}{lccccccccccccc}
\begin{deluxetable}{lcp{0.7cm}cp{1cm}cp{0.8cm}ccp{1.0cm}p{1.0cm}cc}
\tabletypesize{\scriptsize}
\rotate
\tablewidth{0pt}
\tablecaption{Suzaku results of Fe-K features of Sample AGNs.}
\tablehead{
\colhead{(1)} & \colhead{(2)} &  \colhead{(3)} &  \colhead{(4)} &  \colhead{(5)} &  \colhead{(6)} &  \colhead{(7)} &  \colhead{(8)} &  \colhead{(9)} &  \colhead{(10)} &  \colhead{(11)} &  \colhead{(12)} &  \colhead{(13)} \\
\colhead{Object} & \colhead{ObsID} & \colhead{$z$} & \colhead{$\log N_{\rm H}$} & \colhead{$\log F_{\rm X}$,$\log L_{\rm X}$} & \colhead{$E_{64}$}
 & \colhead{$\sigma_{E,64}$} & \colhead{$I_{64}$} & \colhead{$EW_{64}$} & \colhead{$EW_{67}$} & \colhead{$EW_{70}$} &
 \colhead{$\tau_{\rm edge}$} & \colhead{$E_{\rm edge}$} \\ 
\colhead{} & \colhead{} & \colhead{} & \colhead{} & \colhead{} 
& \colhead{(eV)} & \colhead{(eV)} & & \colhead{(eV)} &
 \colhead{(eV)} & \colhead{(eV)} & \colhead{} & \colhead{(keV)}}

\startdata
1H0419-577 & 702041010 & 0.104 & 	\nodata & 	1.50,4.94 & $6364_{ -44}^{+ 39}$ & $<132$ & 	4.5$\pm$ 1.8 & 20$\pm$ 8 & \nodata & \nodata & 	0.12$\pm$0.03 & 7.40$\pm$0.26 \\ 
3C33 & 702059010 & 0.060 & 	$3.74 (0.04)$ & 	1.37,4.29 & $6395_{ -15}^{+ 15}$ & $< 55$ & 	8.2$\pm$ 1.5 & 170$\pm$ 32 & \nodata & \nodata & 	0.63$\pm$0.10 & 7.11$\pm$0.14 \\ 
3C105 & 702074010 & 0.089 & 	$3.61 (0.16)$ & 	1.21,4.48 & $6434_{ -40}^{+ 33}$ & $<108$ & 	7.5$\pm$ 2.7 & 183$\pm$ 65 & \nodata & \nodata & 	0.47$\pm$0.15 & \nodata \\ 
3C120 & 700001030 & 0.033 & 	\nodata & 	1.83,4.23 & $6366_{ -24}^{+ 23}$ & $ 97_{-30}^{+34}$ & 	24.5$\pm$ 4.7 & 50$\pm$ 10 & 7$\pm$ 4 & \nodata & 	0.12$\pm$0.03 & 7.20$\pm$0.27 \\ 
3C273 & 702070010 & 0.158 & 	$<0.47$ & 	2.27,6.09 & \nodata & \nodata & 	\nodata & $<$ 9 & 6$\pm$ 3 & 6$\pm$ 3 & 	\nodata & \nodata \\ 
3C382 & 702125010 & 0.058 & 	\nodata & 	1.75,4.65 & $6415_{ -21}^{+ 20}$ & $ 88_{-26}^{+28}$ & 	16.1$\pm$ 2.9 & 34$\pm$ 6 & \nodata & \nodata & 	0.16$\pm$0.02 & 7.21$\pm$0.15 \\ 
3C390.3 & 701060010 & 0.056 & 	\nodata & 	1.80,4.67 & $6413_{ -19}^{+ 20}$ & $ 84_{-27}^{+28}$ & 	17.1$\pm$ 3.0 & 44$\pm$ 8 & 6$\pm$ 3 & \nodata & 	0.17$\pm$0.03 & 7.29$\pm$0.19 \\ 
3C445 & 702056010 & 0.056 & 	$3.47 (0.05)$ & 	1.51,4.38 & $6385_{ -13}^{+ 12}$ & $ 45_{-32}^{+23}$ & 	12.7$\pm$ 1.9 & 102$\pm$ 15 & \nodata & \nodata & 	0.32$\pm$0.04 & 7.13$\pm$0.13 \\ 
3C452 & 702073010 & 0.081 & 	$3.74 (0.06)$ & 	1.27,4.46 & $6404_{ -21}^{+ 22}$ & $< 77$ & 	6.8$\pm$ 1.8 & 190$\pm$ 49 & \nodata & \nodata & 	0.55$\pm$0.15 & 7.11$\pm$0.28 \\ 
4C+74.26 & 702057010 & 0.104 & 	$<0.50$ & 	1.72,5.15 & $6384_{ -34}^{+ 35}$ & $<134$ & 	13.5$\pm$ 4.3 & 39$\pm$ 13 & \nodata & -14$\pm$ 5 & 	0.28$\pm$0.04 & 7.30$\pm$0.16 \\ 
ARK120 & 702014010 & 0.033 & 	\nodata & 	1.66,4.05 & $6391_{ -15}^{+ 14}$ & $ 72_{-23}^{+25}$ & 	22.3$\pm$ 3.2 & 69$\pm$ 10 & 8$\pm$ 4 & \nodata & 	0.30$\pm$0.03 & 7.11$\pm$0.10 \\ 
ARK564 & 702117010 & 0.025 & 	\nodata & 	1.16,3.30 & \nodata & \nodata & 	\nodata & $<$ 22 & \nodata & \nodata & 	0.12$\pm$0.05 & \nodata \\ 
Centaurus\_A & 100005010 & 0.002$^{\dagger}$ & 	$3.15 (0.02)$ & 	2.78,2.02 & $6399_{ -3}^{+ 3}$ & $ 32_{-7}^{+6}$ & 	242.7$\pm$10.3 & 76$\pm$ 3 & 2$\pm$ 1 & \nodata & 	0.15$\pm$0.01 & 7.13$\pm$0.06 \\ 
Circinus\_Galaxy & 701036010 & 0.001$^{\dagger}$ & 	$4.01 (0.01)$ & 	2.31,1.61 & $6406_{ -1}^{+ 1}$ & $ 33_{-2}^{+2}$ & 	333.3$\pm$ 5.1 & 1379$\pm$ 21 & 143$\pm$ 7 & 71$\pm$ 8 & 	0.62$\pm$0.05 & 7.27$\pm$0.06 \\ 
ESO263-G013 & 702120010 & 0.033 & 	$3.35 (0.05)$ & 	1.27,3.67 & $6406_{ -41}^{+ 36}$ & \nodata & 	5.4$\pm$ 2.7 & 70$\pm$ 35 & \nodata & \nodata & 	0.42$\pm$0.13 & 7.10$\pm$0.26 \\ 
ESO323-G032 & 702119010 & 0.016 & 	$4.07 (0.12)$ & 	0.70,2.45 & $6420_{ -16}^{+ 17}$ & $ 54_{-32}^{+26}$ & 	6.2$\pm$ 1.3 & 823$\pm$ 172 & \nodata & \nodata & 	1.63$\pm$0.77 & \nodata \\ 
ESO506-G027 & 702080010 & 0.025 & 	$3.95 (0.06)$ & 	1.43,3.58 & $6385_{ -11}^{+ 11}$ & $< 50$ & 	20.5$\pm$ 3.3 & 467$\pm$ 75 & \nodata & \nodata & 	1.27$\pm$0.22 & 7.09$\pm$0.13 \\ 
FAIRALL9 & 702043010 & 0.047 & 	\nodata & 	1.57,4.28 & $6384_{ -7}^{+ 7}$ & $ 42_{-18}^{+15}$ & 	25.6$\pm$ 2.3 & 99$\pm$ 9 & 10$\pm$ 4 & \nodata & 	0.24$\pm$0.03 & 7.11$\pm$0.10 \\ 
IC4329A & 702113010 & 0.016 & 	$1.23 (0.11)$ & 	2.28,4.04 & $6373_{ -23}^{+ 22}$ & $ 85_{-31}^{+33}$ & 	59.7$\pm$11.8 & 52$\pm$ 10 & \nodata & \nodata & 	0.18$\pm$0.03 & 7.41$\pm$0.21 \\ 
 & 702113020 & & 	$1.36 (0.07)$ & 	2.38,4.14 & $6382_{ -17}^{+ 17}$ & $< 83$ & 	58.7$\pm$11.6 & 45$\pm$ 9 & \nodata & \nodata & 	0.19$\pm$0.03 & 7.18$\pm$0.14 \\ 
 & 702113030 & & 	$1.31 (0.09)$ & 	2.36,4.12 & $6369_{ -18}^{+ 17}$ & $ 65_{-31}^{+29}$ & 	66.6$\pm$11.9 & 53$\pm$ 9 & 12$\pm$ 4 & -7$\pm$ 4 & 	0.12$\pm$0.03 & 7.44$\pm$0.28 \\ 
 & 702113040 & & 	$1.25 (0.11)$ & 	2.33,4.09 & $6377_{ -18}^{+ 18}$ & $ 67_{-28}^{+27}$ & 	66.1$\pm$12.1 & 59$\pm$ 11 & 8$\pm$ 5 & \nodata & 	0.16$\pm$0.03 & 7.26$\pm$0.20 \\ 
 & 702113050 & & 	$0.94 (0.28)$ & 	2.20,3.96 & $6365_{ -11}^{+ 17}$ & $< 54$ & 	51.4$\pm$ 9.9 & 69$\pm$ 13 & 9$\pm$ 6 & \nodata & 	0.25$\pm$0.04 & 7.31$\pm$0.18 \\ 
IGRJ16185-5928 & 702123010 & 0.035 & 	\nodata & 	1.12,3.57 & $6431_{ -27}^{+ 30}$ & $< 87$ & 	6.2$\pm$ 1.9 & 72$\pm$ 22 & \nodata & \nodata & 	0.22$\pm$0.07 & \nodata \\ 
IGRJ21247+5058 & 702027010 & 0.020 & 	$3.32 (0.05)$ & 	2.22,4.18 & $6398_{ -39}^{+ 38}$ & $ 97_{-46}^{+67}$ & 	18.5$\pm$ 5.9 & 21$\pm$ 7 & \nodata & \nodata & 	0.12$\pm$0.02 & 7.24$\pm$0.19 \\ 
IRAS18325-5926 & 702118010 & 0.020 & 	$2.03 (0.74)$ & 	1.44,3.41 & $6374_{ -26}^{+ 26}$ & $< 88$ & 	9.6$\pm$ 2.9 & 38$\pm$ 11 & 23$\pm$ 5 & \nodata & 	0.21$\pm$0.04 & 7.12$\pm$0.19 \\ 
IRAS19254-7245 & 701052010 & 0.062 & 	$4.57 (0.16)$ & 	1.08,4.01 & \nodata & \nodata & 	1.0$\pm$ 0.6 & 515$\pm$ 275 & 540$\pm$168 & 286$\pm$137 & 	\nodata & \nodata \\ 
M81 & 701022010 & 0.000$^{\dagger}$ & 	$1.09 (0.24)$ & 	1.20,0.36 & \nodata & \nodata & 	5.2$\pm$ 2.1 & 45$\pm$ 18 & 37$\pm$ 9 & \nodata & 	0.13$\pm$0.06 & \nodata \\ 
M106 & 701095010 & 0.002$^{\dagger}$ & 	$3.17 (0.11)$ & 	1.23,1.08 & $6417_{ -32}^{+ 33}$ & $< 91$ & 	5.2$\pm$ 1.9 & 43$\pm$ 15 & \nodata & \nodata & 	0.11$\pm$0.04 & \nodata \\ 
MCG+04-48-002 & 702081010 & 0.014 & 	$3.83 (0.07)$ & 	1.57,3.20 & $6385_{ -25}^{+ 21}$ & $< 64$ & 	8.7$\pm$ 2.7 & 184$\pm$ 58 & \nodata & \nodata & 	0.89$\pm$0.14 & 7.09$\pm$0.18 \\ 
MCG+8-11-11 & 702112010 & 0.021 & 	$0.81 (0.18)$ & 	2.06,4.04 & $6379_{ -7}^{+ 7}$ & $ 61_{-12}^{+12}$ & 	54.3$\pm$ 4.4 & 77$\pm$ 6 & 4$\pm$ 3 & \nodata & 	0.12$\pm$0.02 & 7.11$\pm$0.14 \\ 
MCG-5-23-16 & 700002010 & 0.009 & 	$3.51 (0.04)$ & 	2.25,3.45 & $6390_{ -7}^{+ 7}$ & $ 68_{-11}^{+11}$ & 	80.2$\pm$ 5.5 & 76$\pm$ 5 & \nodata & 4$\pm$ 2 & 	0.18$\pm$0.02 & 7.20$\pm$0.09 \\ 
MCG-6-30-15 & 700007010 & 0.008 & 	$1.34 (0.05)$ & 	1.78,2.90 & $6469_{ -25}^{+ 24}$ & $171_{-29}^{+28}$ & 	18.2$\pm$ 2.7 & 40$\pm$ 6 & -12$\pm$ 3 & -19$\pm$ 3 & 	0.27$\pm$0.01 & 7.14$\pm$0.07 \\ 
 & 700007020 & & 	$1.34 (0.07)$ & 	1.81,2.93 & $6414_{ -27}^{+ 26}$ & $149_{-32}^{+34}$ & 	18.4$\pm$ 3.2 & 44$\pm$ 8 & -12$\pm$ 3 & -21$\pm$ 3 & 	0.28$\pm$0.03 & 7.31$\pm$0.10 \\ 
 & 700007030 & & 	$1.34 (0.06)$ & 	1.83,2.95 & $6400_{ -23}^{+ 23}$ & $102_{-33}^{+34}$ & 	17.3$\pm$ 3.2 & 38$\pm$ 7 & -10$\pm$ 3 & -19$\pm$ 3 & 	0.29$\pm$0.02 & 7.28$\pm$0.08 \\ 
MRK3 & 100040010 & 0.013 & 	$3.88 (0.01)$ & 	1.95,3.56 & $6408_{ -3}^{+ 3}$ & $ 28_{-10}^{+7}$ & 	51.1$\pm$ 2.4 & 454$\pm$ 21 & 56$\pm$ 8 & 15$\pm$ 8 & 	0.76$\pm$0.05 & 7.13$\pm$0.05 \\ 
MRK79 & 702044010 & 0.022 & 	$0.98 (0.32)$ & 	1.40,3.45 & $6397_{ -10}^{+ 10}$ & $ 44_{-26}^{+19}$ & 	21.6$\pm$ 2.7 & 130$\pm$ 16 & 15$\pm$ 7 & \nodata & 	0.24$\pm$0.05 & 7.18$\pm$0.18 \\ 
MRK110 & 702124010 & 0.035 & 	\nodata & 	1.52,3.98 & $6404_{ -26}^{+ 27}$ & $ 70_{-38}^{+40}$ & 	10.3$\pm$ 2.6 & 44$\pm$ 11 & \nodata & 10$\pm$ 5 & 	0.10$\pm$0.04 & \nodata \\ 
MRK273 & 701050010 & 0.038 & 	$4.09 (0.13)$ & 	0.93,3.43 & $6384_{ -26}^{+ 19}$ & $< 64$ & 	3.2$\pm$ 1.1 & 765$\pm$ 258 & 234$\pm$149 & \nodata & 	1.07$\pm$1.01 & \nodata \\ 
MRK335 & 701031010 & 0.026 & 	\nodata & 	1.29,3.48 & $6400_{ -25}^{+ 27}$ & $ 89_{-31}^{+33}$ & 	8.0$\pm$ 1.7 & 49$\pm$ 11 & 16$\pm$ 5 & \nodata & 	0.33$\pm$0.03 & 7.16$\pm$0.11 \\ 
MRK359 & 701082010 & 0.017 & 	\nodata & 	0.92,2.75 & $6380_{ -24}^{+ 22}$ & $ 81_{-35}^{+33}$ & 	6.3$\pm$ 1.3 & 122$\pm$ 25 & 24$\pm$ 11 & \nodata & 	0.25$\pm$0.08 & \nodata \\ 
MRK417 & 702078010 & 0.033 & 	$3.66 (0.09)$ & 	1.49,3.87 & $6371_{ -31}^{+ 25}$ & $< 88$ & 	7.4$\pm$ 2.7 & 126$\pm$ 46 & \nodata & \nodata & 	0.61$\pm$0.13 & 7.08$\pm$0.19 \\ 
MRK509 & 701093010 & 0.034 & 	\nodata & 	1.89,4.33 & $6411_{ -31}^{+ 31}$ & $<112$ & 	29.1$\pm$ 9.2 & 55$\pm$ 17 & 14$\pm$ 8 & \nodata & 	0.12$\pm$0.05 & \nodata \\ 
 & 701093020 & & 	\nodata & 	1.91,4.34 & $6407_{ -40}^{+ 38}$ & $ 89_{-62}^{+58}$ & 	23.2$\pm$ 7.0 & 42$\pm$ 13 & 12$\pm$ 6 & \nodata & 	0.13$\pm$0.04 & \nodata \\ 
 & 701093030 & & 	\nodata & 	1.86,4.30 & $6404_{ -43}^{+ 41}$ & $<170$ & 	31.7$\pm$ 9.9 & 63$\pm$ 20 & 21$\pm$ 9 & \nodata & 	0.07$\pm$0.06 & \nodata \\ 
 & 701093040 & & 	\nodata & 	1.86,4.30 & $6396_{ -25}^{+ 32}$ & $< 78$ & 	24.1$\pm$ 7.4 & 50$\pm$ 15 & 17$\pm$ 7 & \nodata & 	0.10$\pm$0.05 & \nodata \\ 
MRK766 & 701035010 & 0.013 & 	$0.60 (0.62)$ & 	1.26,2.84 & $6450_{ -41}^{+ 40}$ & $<143$ & 	5.7$\pm$ 2.1 & 42$\pm$ 16 & 26$\pm$ 8 & -28$\pm$ 7 & 	0.27$\pm$0.04 & 7.13$\pm$0.19 \\ 
 & 701035020 & & 	$1.75 (0.06)$ & 	1.28,2.85 & $6344_{ -44}^{+ 48}$ & $148_{-69}^{+51}$ & 	8.2$\pm$ 2.4 & 59$\pm$ 18 & -39$\pm$ 7 & -29$\pm$ 7 & 	0.13$\pm$0.05 & \nodata \\ 
MRK841 & 701084010 & 0.036 & 	\nodata & 	1.44,3.93 & $6375_{ -25}^{+ 24}$ & $< 55$ & 	10.1$\pm$ 3.3 & 65$\pm$ 22 & 21$\pm$ 10 & \nodata & 	0.23$\pm$0.07 & 7.15$\pm$0.26 \\ 
 & 701084020 & & 	\nodata & 	1.50,3.98 & $6326_{ -26}^{+ 57}$ & $184_{-63}^{+15}$ & 	10.9$\pm$ 3.5 & 66$\pm$ 21 & \nodata & \nodata & 	0.21$\pm$0.07 & \nodata \\ 
MRK1073 & 701007020 & 0.023 & 	$4.37 (0.32)$ & 	0.82,2.90 & $6408_{ -39}^{+ 47}$ & $<112$ & 	3.5$\pm$ 1.6 & 616$\pm$ 287 & \nodata & \nodata & 	\nodata & \nodata \\ 
MRK1210 & 702111010 & 0.013 & 	$3.54 (0.08)$ & 	1.60,3.21 & $6386_{ -14}^{+ 13}$ & $< 70$ & 	16.2$\pm$ 3.1 & 187$\pm$ 35 & -30$\pm$ 15 & -35$\pm$ 15 & 	0.77$\pm$0.09 & 7.11$\pm$0.13 \\ 
MRK1239 & 702031010 & 0.020 & 	$3.66 (0.11)$ & 	0.88,2.83 & $6425_{ -33}^{+ 45}$ & $<110$ & 	3.1$\pm$ 1.3 & 179$\pm$ 72 & \nodata & \nodata & 	1.61$\pm$0.39 & 7.15$\pm$0.20 \\ 
NGC1052 & 702058010 & 0.005 & 	$3.40 (0.07)$ & 	1.25,1.99 & $6383_{ -19}^{+ 20}$ & $< 78$ & 	8.8$\pm$ 2.2 & 102$\pm$ 25 & \nodata & -17$\pm$ 11 & 	0.29$\pm$0.07 & 7.13$\pm$0.24 \\ 
NGC1068 & 701039010 & 0.004 & 	$3.99 (0.02)$ & 	1.39,1.89 & $6413_{ -9}^{+ 9}$ & $ 61_{-16}^{+14}$ & 	52.0$\pm$ 5.0 & 706$\pm$ 68 & 290$\pm$ 32 & \nodata & 	0.71$\pm$0.18 & 7.11$\pm$0.19 \\ 
NGC1142 & 702079010 & 0.029 & 	$3.80 (0.08)$ & 	1.53,3.79 & $6387_{ -16}^{+ 15}$ & $ 61_{-22}^{+21}$ & 	17.4$\pm$ 3.1 & 361$\pm$ 65 & \nodata & \nodata & 	0.63$\pm$0.17 & 7.11$\pm$0.21 \\ 
NGC1365 & 702047010 & 0.005$^{\dagger}$ & 	$3.01 (0.11)$ & 	1.77,2.33 & $6419_{ -19}^{+ 18}$ & $137_{-15}^{+17}$ & 	15.6$\pm$ 2.1 & 87$\pm$ 12 & -90$\pm$ 4 & -97$\pm$ 4 & 	0.28$\pm$0.03 & 7.11$\pm$0.10 \\ 
NGC1386 & 702002010 & 0.003 & 	$4.31 (0.08)$ & 	0.98,1.44 & \nodata & \nodata & 	5.7$\pm$ 2.0 & 996$\pm$ 342 & \nodata & \nodata & 	2.42$\pm$1.71 & \nodata \\ 
NGC2110 & 100024010 & 0.008 & 	$3.15 (0.04)$ & 	2.41,3.54 & $6403_{ -6}^{+ 6}$ & $ 45_{-13}^{+11}$ & 	71.9$\pm$ 5.3 & 50$\pm$ 4 & \nodata & \nodata & 	0.10$\pm$0.01 & 7.21$\pm$0.11 \\ 
NGC2273 & 702003010 & 0.006 & 	$3.93 (0.06)$ & 	1.02,1.94 & $6390_{ -8}^{+ 7}$ & $< 21$ & 	22.2$\pm$ 2.0 & 1864$\pm$ 168 & \nodata & \nodata & 	\nodata & \nodata \\ 
NGC3079 & 803039020 & 0.004 & 	$4.33 (0.06)$ & 	1.31,1.79 & $6407_{ -16}^{+ 23}$ & $< 67$ & 	3.9$\pm$ 1.2 & 545$\pm$ 160 & \nodata & -155$\pm$100 & 	0.90$\pm$0.72 & \nodata \\ 
NGC3227 & 703022010 & 0.004 & 	$1.65 (0.04)$ & 	2.01,2.53 & $6381_{ -10}^{+ 10}$ & $ 49_{-25}^{+19}$ & 	37.2$\pm$ 4.3 & 81$\pm$ 9 & -24$\pm$ 4 & -26$\pm$ 4 & 	0.22$\pm$0.03 & 7.31$\pm$0.13 \\ 
 & 703022020 & & 	$1.99 (0.04)$ & 	1.91,2.43 & $6395_{ -11}^{+ 11}$ & $ 72_{-17}^{+17}$ & 	42.2$\pm$ 4.5 & 164$\pm$ 17 & \nodata & -14$\pm$ 7 & 	0.38$\pm$0.05 & 7.24$\pm$0.12 \\ 
 & 703022030 & & 	$2.28 (0.02)$ & 	1.90,2.42 & $6387_{ -12}^{+ 12}$ & $ 50_{-37}^{+24}$ & 	38.2$\pm$ 5.0 & 113$\pm$ 15 & -24$\pm$ 6 & -32$\pm$ 6 & 	0.40$\pm$0.05 & 7.17$\pm$0.10 \\ 
 & 703022040 & & 	$1.06 (0.47)$ & 	1.75,2.27 & $6394_{ -9}^{+ 9}$ & $ 44_{-22}^{+17}$ & 	37.1$\pm$ 3.8 & 278$\pm$ 28 & 36$\pm$ 11 & \nodata & 	0.55$\pm$0.08 & 7.13$\pm$0.12 \\ 
 & 703022050 & & 	$2.38 (0.02)$ & 	1.83,2.36 & $6395_{ -13}^{+ 13}$ & $< 70$ & 	34.3$\pm$ 5.1 & 115$\pm$ 17 & -17$\pm$ 7 & -47$\pm$ 7 & 	0.32$\pm$0.05 & 7.11$\pm$0.14 \\ 
 & 703022060 & & 	$2.31 (0.03)$ & 	1.74,2.27 & $6400_{ -15}^{+ 8}$ & $< 44$ & 	31.6$\pm$ 4.5 & 143$\pm$ 20 & -14$\pm$ 8 & -17$\pm$ 8 & 	0.40$\pm$0.06 & 7.20$\pm$0.13 \\ 
NGC3281 & 703033010 & 0.011 & 	$3.79 (0.06)$ & 	1.82,3.22 & $6403_{ -10}^{+ 10}$ & $< 60$ & 	26.8$\pm$ 3.6 & 400$\pm$ 54 & -33$\pm$ 21 & -40$\pm$ 23 & 	1.11$\pm$0.16 & 7.23$\pm$0.13 \\ 
NGC3393 & 702004010 & 0.013 & 	$4.00 (0.10)$ & 	1.10,2.64 & $6407_{ -24}^{+ 32}$ & $< 52$ & 	4.3$\pm$ 1.8 & 507$\pm$ 214 & \nodata & \nodata & 	1.47$\pm$0.88 & \nodata \\ 
NGC3516 & 100031010 & 0.009 & 	$3.26 (0.04)$ & 	1.91,3.14 & $6401_{ -4}^{+ 4}$ & $ 58_{-7}^{+7}$ & 	53.2$\pm$ 2.9 & 146$\pm$ 8 & -19$\pm$ 3 & -18$\pm$ 3 & 	0.36$\pm$0.02 & 7.27$\pm$0.06 \\ 
NGC3783 & 701033010 & 0.010 & 	$1.57 (0.04)$ & 	2.01,3.33 & $6397_{ -6}^{+ 6}$ & $ 54_{-12}^{+10}$ & 	69.2$\pm$ 4.6 & 125$\pm$ 8 & -12$\pm$ 3 & \nodata & 	0.23$\pm$0.02 & 7.28$\pm$0.10 \\ 
NGC4051 & 703023010 & 0.002$^{\dagger}$ & 	$<-1.59$ & 	1.56,2.07 & $6405_{ -10}^{+ 10}$ & $< 59$ & 	16.2$\pm$ 1.9 & 70$\pm$ 8 & \nodata & -20$\pm$ 3 & 	0.33$\pm$0.03 & 7.14$\pm$0.07 \\ 
 & 700004010 & & 	$1.27 (0.17)$ & 	1.32,1.84 & $6395_{ -11}^{+ 11}$ & $ 51_{-21}^{+19}$ & 	13.9$\pm$ 1.8 & 133$\pm$ 17 & \nodata & -29$\pm$ 7 & 	0.32$\pm$0.05 & 7.11$\pm$0.14 \\ 
 & 703023020 & & 	\nodata & 	1.46,1.97 & $6420_{ -19}^{+ 18}$ & $< 95$ & 	15.8$\pm$ 2.9 & 91$\pm$ 17 & \nodata & -24$\pm$ 7 & 	0.30$\pm$0.05 & 7.29$\pm$0.18 \\ 
NGC4151 & 701034010 & 0.003$^{\dagger}$ & 	$3.39 (0.02)$ & 	2.28,2.93 & $6392_{ -2}^{+ 2}$ & $ 35_{-5}^{+4}$ & 	166.1$\pm$ 5.1 & 258$\pm$ 8 & \nodata & -5$\pm$ 3 & 	0.38$\pm$0.02 & 7.14$\pm$0.04 \\ 
NGC4388 & 800017010 & 0.008 & 	$3.55 (0.01)$ & 	2.13,3.32 & $6405_{ -2}^{+ 2}$ & $ 38_{-6}^{+5}$ & 	82.5$\pm$ 3.0 & 232$\pm$ 8 & -9$\pm$ 3 & \nodata & 	0.47$\pm$0.02 & 7.15$\pm$0.04 \\ 
NGC4395 & 702001010 & 0.001 & 	$1.63 (0.14)$ & 	1.17,0.59 & $6382_{ -36}^{+ 36}$ & $<103$ & 	4.0$\pm$ 1.5 & 70$\pm$ 27 & \nodata & \nodata & 	0.19$\pm$0.08 & \nodata \\ 
NGC4507 & 702048010 & 0.012 & 	$3.78 (0.02)$ & 	1.87,3.36 & $6405_{ -3}^{+ 3}$ & $ 39_{-8}^{+7}$ & 	53.1$\pm$ 2.6 & 603$\pm$ 29 & 55$\pm$ 10 & 39$\pm$ 10 & 	0.78$\pm$0.07 & 7.28$\pm$0.07 \\ 
NGC4593 & 702040010 & 0.009 & 	\nodata & 	1.42,2.67 & $6420_{ -6}^{+ 6}$ & $ 39_{-15}^{+12}$ & 	25.8$\pm$ 2.1 & 244$\pm$ 20 & \nodata & \nodata & 	0.27$\pm$0.05 & 7.24$\pm$0.17 \\ 
NGC4945 & 100008030 & 0.002$^{\dagger}$ & 	$4.40 (0.02)$ & 	2.17,1.49 & $6409_{ -4}^{+ 4}$ & $ 39_{-11}^{+9}$ & 	32.4$\pm$ 2.3 & 638$\pm$ 45 & 153$\pm$ 21 & \nodata & 	0.23$\pm$0.11 & 7.15$\pm$0.27 \\ 
NGC4968 & 701005010 & 0.010 & 	$<3.64$ & 	0.04,1.38 & $6430_{ -24}^{+ 16}$ & $< 68$ & 	5.6$\pm$ 1.7 & 1798$\pm$ 565 & \nodata & \nodata & 	\nodata & \nodata \\ 
NGC4992 & 701080010 & 0.025 & 	$3.68 (0.09)$ & 	1.51,3.66 & $6387_{ -24}^{+ 23}$ & $110_{-29}^{+30}$ & 	13.7$\pm$ 2.7 & 281$\pm$ 55 & \nodata & \nodata & 	0.75$\pm$0.20 & 7.11$\pm$0.20 \\ 
NGC5135 & 702005010 & 0.014 & 	$4.33 (0.28)$ & 	1.10,2.71 & $6381_{ -15}^{+ 14}$ & $< 53$ & 	8.2$\pm$ 2.0 & 1560$\pm$ 382 & \nodata & \nodata & 	\nodata & \nodata \\ 
NGC5506 & 701030010 & 0.006 & 	$2.44 (0.07)$ & 	2.30,3.23 & $6420_{ -10}^{+ 10}$ & $ 61_{-18}^{+16}$ & 	82.3$\pm$ 8.4 & 64$\pm$ 7 & 14$\pm$ 3 & 10$\pm$ 3 & 	0.21$\pm$0.02 & 7.24$\pm$0.10 \\ 
 & 701030020 & & 	$2.55 (0.07)$ & 	2.33,3.26 & $6403_{ -12}^{+ 12}$ & $106_{-16}^{+17}$ & 	85.6$\pm$ 7.9 & 64$\pm$ 6 & 20$\pm$ 3 & 9$\pm$ 3 & 	0.23$\pm$0.02 & 7.26$\pm$0.09 \\ 
 & 701030030 & & 	$2.69 (0.14)$ & 	2.32,3.25 & $6386_{ -11}^{+ 11}$ & $ 79_{-15}^{+16}$ & 	74.8$\pm$ 8.0 & 64$\pm$ 7 & 5$\pm$ 3 & \nodata & 	0.27$\pm$0.02 & 7.19$\pm$0.09 \\ 
NGC5548 & 702042010 & 0.017 & 	\nodata & 	1.50,3.32 & $6395_{ -14}^{+ 11}$ & $< 32$ & 	18.3$\pm$ 3.7 & 204$\pm$ 41 & \nodata & 35$\pm$ 17 & 	\nodata & \nodata \\ 
 & 702042020 & & 	$1.00 (0.49)$ & 	1.64,3.45 & $6367_{ -18}^{+ 17}$ & $ 60_{-31}^{+28}$ & 	20.6$\pm$ 3.8 & 137$\pm$ 25 & \nodata & \nodata & 	0.13$\pm$0.07 & \nodata \\ 
 & 702042040 & & 	$0.85 (0.53)$ & 	1.64,3.46 & $6368_{ -23}^{+ 23}$ & $< 78$ & 	18.6$\pm$ 5.2 & 68$\pm$ 19 & -14$\pm$ 8 & \nodata & 	0.10$\pm$0.06 & \nodata \\ 
 & 702042050 & & 	$<0.68$ & 	1.66,3.48 & $6380_{ -16}^{+ 15}$ & $< 56$ & 	22.7$\pm$ 4.8 & 120$\pm$ 26 & \nodata & \nodata & 	0.10$\pm$0.07 & \nodata \\ 
 & 702042060 & & 	$1.05 (0.32)$ & 	1.82,3.64 & $6393_{ -37}^{+ 37}$ & $<133$ & 	18.5$\pm$ 6.3 & 52$\pm$ 18 & -18$\pm$ 7 & \nodata & 	0.07$\pm$0.06 & \nodata \\ 
 & 702042070 & & 	$<0.63$ & 	1.73,3.55 & $6363_{ -15}^{+ 15}$ & $< 53$ & 	24.9$\pm$ 5.3 & 108$\pm$ 23 & \nodata & \nodata & 	0.10$\pm$0.07 & \nodata \\ 
 & 702042080 & & 	$0.98 (0.60)$ & 	1.60,3.42 & $6386_{ -16}^{+ 15}$ & $< 67$ & 	20.6$\pm$ 4.0 & 157$\pm$ 31 & \nodata & \nodata & 	0.34$\pm$0.09 & 7.13$\pm$0.24 \\ 
NGC5728 & 701079010 & 0.009 & 	$4.04 (0.05)$ & 	1.64,2.92 & $6396_{ -9}^{+ 9}$ & $< 47$ & 	20.6$\pm$ 2.5 & 850$\pm$ 102 & \nodata & \nodata & 	1.59$\pm$0.29 & 7.08$\pm$0.12 \\ 
NGC6300 & 702049010 & 0.004 & 	$3.34 (0.04)$ & 	1.83,2.31 & $6363_{ -12}^{+ 12}$ & $< 64$ & 	22.2$\pm$ 2.8 & 84$\pm$ 11 & -17$\pm$ 4 & -27$\pm$ 4 & 	0.43$\pm$0.03 & 7.11$\pm$0.07 \\ 
NGC7172 & 703030010 & 0.009 & 	$3.12 (0.06)$ & 	2.09,3.31 & $6413_{ -16}^{+ 16}$ & $ 74_{-27}^{+26}$ & 	33.7$\pm$ 4.9 & 50$\pm$ 7 & -13$\pm$ 3 & -16$\pm$ 3 & 	0.29$\pm$0.02 & 7.25$\pm$0.08 \\ 
NGC7213 & 701029010 & 0.006 & 	\nodata & 	1.59,2.46 & $6402_{ -8}^{+ 8}$ & $< 45$ & 	20.1$\pm$ 2.4 & 83$\pm$ 10 & 18$\pm$ 4 & 17$\pm$ 4 & 	0.13$\pm$0.03 & 7.17$\pm$0.22 \\ 
NGC7314 & 702015010 & 0.005 & 	$<2.99$ & 	1.19,1.90 & $6393_{ -14}^{+ 14}$ & $ 49_{-23}^{+20}$ & 	11.8$\pm$ 2.0 & 124$\pm$ 21 & \nodata & 29$\pm$ 9 & 	0.10$\pm$0.06 & \nodata \\ 
NGC7582 & 702052010 & 0.005 & 	$3.53 (0.11)$ & 	1.74,2.51 & $6409_{ -11}^{+ 19}$ & $< 48$ & 	22.7$\pm$ 5.4 & 221$\pm$ 52 & \nodata & \nodata & 	0.57$\pm$0.14 & 7.20$\pm$0.23 \\ 
 & 702052020 & & 	$3.69 (0.07)$ & 	1.66,2.44 & $6385_{ -15}^{+ 16}$ & $< 64$ & 	23.7$\pm$ 4.8 & 292$\pm$ 59 & 45$\pm$ 27 & \nodata & 	0.70$\pm$0.16 & 7.19$\pm$0.21 \\ 
 & 702052030 & & 	$3.79 (0.05)$ & 	1.67,2.46 & $6412_{ -13}^{+ 13}$ & $< 60$ & 	18.9$\pm$ 3.3 & 394$\pm$ 69 & \nodata & \nodata & 	0.72$\pm$0.19 & 7.23$\pm$0.22 \\ 
 & 702052040 & & 	$3.85 (0.05)$ & 	1.58,2.37 & $6409_{ -12}^{+ 12}$ & $ 46_{-25}^{+19}$ & 	19.7$\pm$ 3.1 & 454$\pm$ 72 & \nodata & \nodata & 	0.72$\pm$0.19 & 7.15$\pm$0.30 \\ 
PDS456 & 701056010 & 0.184 & 	\nodata & 	0.62,4.58 & \nodata & \nodata & 	\nodata & $<$ 17 & \nodata & \nodata & 	\nodata & \nodata \\ 
PKS2356-61 & 801016010 & 0.096 & 	$3.41 (0.12)$ & 	0.96,4.33 & \nodata & \nodata & 	\nodata & $<$ 42 & \nodata & \nodata & 	0.21$\pm$0.09 & \nodata \\ 
RBS1124 & 702114010 & 0.208 & 	\nodata & 	1.02,5.10 & $6412_{-112}^{+ 87}$ & $<199$ & 	\nodata & $<$ 59 & \nodata & \nodata & 	0.10$\pm$0.07 & \nodata \\ 
SJ0134.1-3625$^{\star}$ & 703016010 & 0.030 & 	$4.31 (0.17)$ & 	1.26,2.60 & $6300_{ 0}^{+ 37}$ & \nodata & 	\nodata & $<$ 224 & \nodata & \nodata & 	1.93$\pm$1.75 & \nodata \\ 
SJ0138.6-4001$^{\star}$ & 701015010 & 0.025 & 	$3.75 (0.09)$ & 	1.57,3.72 & $6386_{ -28}^{+ 26}$ & $< 90$ & 	15.4$\pm$ 4.8 & 222$\pm$ 69 & \nodata & \nodata & 	0.89$\pm$0.19 & 7.11$\pm$0.24 \\ 
SJ0255.2-0011$^{\star}$ & 701013010 & 0.029 & 	$3.76 (0.04)$ & 	1.72,3.99 & $6388_{ -9}^{+ 9}$ & $ 38_{-19}^{+15}$ & 	17.5$\pm$ 2.3 & 221$\pm$ 29 & \nodata & \nodata & 	0.86$\pm$0.07 & 7.10$\pm$0.09 \\ 
SJ0318.7+6828$^{\star}$ & 702075010 & 0.090 & 	$3.18 (0.26)$ & 	1.01,4.32 & $6330_{ -30}^{+ 31}$ & $< 75$ & 	3.8$\pm$ 1.7 & 52$\pm$ 23 & \nodata & \nodata & 	0.13$\pm$0.08 & \nodata \\ 
SJ0501.9-3239$^{\star}$ & 703014010 & 0.010 & 	\nodata & 	1.77,3.12 & \nodata & \nodata & 	23.6$\pm$ 5.6 & 72$\pm$ 17 & \nodata & \nodata & 	0.11$\pm$0.05 & \nodata \\ 
SJ0505.7-2348$^{\star}$ & 701014010 & 0.041 & 	$3.21 (0.14)$ & 	1.51,4.10 & $6420_{ -18}^{+ 26}$ & $< 55$ & 	9.8$\pm$ 2.9 & 57$\pm$ 17 & \nodata & 16$\pm$ 8 & 	0.14$\pm$0.05 & \nodata \\ 
SJ0601.9-8636$^{\star}$ & 701018010 & 0.006 & 	$3.86 (0.12)$ & 	1.09,2.02 & $6399_{ -15}^{+ 9}$ & $< 47$ & 	21.6$\pm$ 4.1 & 1254$\pm$ 239 & \nodata & 179$\pm$105 & 	1.43$\pm$0.74 & 7.09$\pm$0.28 \\ 
SJ0959.5-2258$^{\star}$ & 703013010 & 0.010 & 	$3.89 (0.06)$ & 	1.52,2.86 & $6424_{ -15}^{+ 15}$ & $< 64$ & 	15.6$\pm$ 3.0 & 270$\pm$ 52 & -43$\pm$ 22 & -50$\pm$ 23 & 	1.09$\pm$0.14 & 7.21$\pm$0.13 \\ 
SJ1200.8+0650$^{\star}$ & 703009010 & 0.036 & 	$3.29 (0.12)$ & 	1.44,3.92 & $6339_{ -39}^{+ 43}$ & $112_{-41}^{+45}$ & 	6.9$\pm$ 2.3 & 55$\pm$ 18 & \nodata & \nodata & 	0.35$\pm$0.07 & 7.30$\pm$0.19 \\ 
SJ1628.1+5145$^{\star}$ & 701016010 & 0.055 & 	$3.25 (0.20)$ & 	1.60,4.44 & $6429_{-110}^{+ 45}$ & \nodata & 	7.8$\pm$ 5.1 & 45$\pm$ 30 & -33$\pm$ 15 & \nodata & 	0.30$\pm$0.10 & \nodata \\ 
SJ2009.0-6103$^{\star}$ & 703015010 & 0.015 & 	$1.05 (0.38)$ & 	1.63,2.98 & $6346_{ -46}^{+ 94}$ & \nodata & 	9.8$\pm$ 4.9 & 44$\pm$ 22 & \nodata & \nodata & 	0.25$\pm$0.07 & \nodata \\ 
TON\_S180 & 701021010 & 0.062 & 	\nodata & 	0.94,3.87 & \nodata & \nodata & 	\nodata & $<$ 21 & 15$\pm$ 9 & \nodata & 	0.17$\pm$0.07 & \nodata \\

\enddata
\tablenotetext{a}{Swift/BAT sources. The name of Swift J*** is reduced to SJ***.}
\tablenotetext{b}{The distance of these objects in unit of
 Mpc are as follows in parentheses: Centaurus\_A (3.9), Circinus\_Galaxy
 (4.2), M81 (3.6), M106 (7.9), NGC 1365 (18), NGC 4051 (17), NGC 4151
 (20), NGC 4945 (4.3).}
\tablecomments{(4): Logarithmic absorption Hydrogen column density 
at the rest frame of the object in unit of $10^{20}$ cm$^{-2}$. Errors
 in parentheses represent the relative error. (5):
Logarithmic observed flux in 10--50 keV in unit of $10^{-12}$ erg
cm$^{-2}$ s$^{-1}$, and logarithmic X-ray luminosity in 15--50 keV
in unit of $10^{40}$ erg s$^{-1}$. (6)--(8): Center energy, width, and
intensity of the 6.4 keV line. The intensity is in unit of $10^{-6}$ ph
 cm$^{-2}$ s$^{-1}$. (9)--(11): Equivalent width of the 
6.4, 6.7, and 7.0 keV lines. Negative values represent the absorption line.
(12)--(13): Depth and energy of the Fe-K edge.}
\end{deluxetable}

\clearpage

\end{document}